\documentclass[12pt, twocolumn]{aastex63}

\usepackage[english]{babel}
\usepackage{amsmath,amssymb,amsfonts, bm,bbm,slashed, subdepth}
\usepackage{mathalfa}
\usepackage{graphicx}
\usepackage{hyperref}
\usepackage{enumerate}
\usepackage{epsfig, subfigure}
\usepackage{multirow}

\usepackage{braket}

\allowdisplaybreaks

\DeclareMathAlphabet\mathbfcal{OMS}{cmsy}{b}{n}

\DeclareMathOperator{\Tr}{Tr}

\let\oldhat\hat
\renewcommand{\vec}[1]{\mathbf{#1}}
\renewcommand{\hat}[1]{\bm\oldhat{\mathbf{#1}}}
\newcommand{\thinhat}[1]{\oldhat{{#1}}}

\newcommand{\lss}[1]{\textcolor{black}{#1}}

\newcommand{\Eeff}{\varepsilon^*\sigma(v_i)}
\newcommand{\EeffQD}[1]{\varepsilon^*\sigma_{#1}(v_i)}

\newcommand{\Eeffclass}{\varepsilon^*\sigma}

\newcommand{\singlet}{{\uparrow \downarrow}}
\newcommand{\triplet}{{\uparrow \uparrow}}

\newcommand{\Fconf}{{}_1F_1}
\newcommand{\WFnorm}{B}

\renewcommand{\Im}{\operatorname{Im}}
\renewcommand{\Re}{\operatorname{Re}}

	\usepackage[]{tikz}

\begin{document}

\title{Exact Theory of Non-relativistic Quadrupole Bremsstrahlung}
\author[0000-0002-9165-4813]{Josef Pradler}

\affiliation{Institute of High Energy Physics, Austrian Academy of Sciences, Nikolsdorfergasse 18, 1050 Vienna, Austria}
\email{josef.pradler@oeaw.ac.at}
\author[0000-0002-7059-2094]{Lukas Semmelrock}

\affiliation{Institute of High Energy Physics, Austrian Academy of Sciences, Nikolsdorfergasse 18, 1050 Vienna, Austria}
\email{lukas.semmelrock@oeaw.ac.at}

\begin{abstract}
The quantum mechanical treatment of quadrupole radiation due to bremsstrahlung exact to all orders in the Coulomb interaction of two non-relativistically colliding charged spin-0 and unpolarized  spin-1/2 particles is presented. We calculate the elements of the quadrupole tensor, and present analytical solutions for the double differential photon emission cross section of identical and non-identical particles. Contact is made to Born-level and quasi-classical results in the respective kinematic limits and effective energy loss rates are obtained. A generally valid formula for soft-photon emission is established and an approximate formula across the entire kinematic regime is constructed. The results apply to bremsstrahlung emission in the scattering of pairs of electrons and are hence relevant in the study of astrophysical phenomena. A definition of a Gaunt factor is proposed which should facilitate broad applicability of the results.
\vspace{30pt}
\end{abstract}

\section{Introduction}
The emission of a photon in the scattering of two electrically charged particles is one of the most fundamental processes of quantum electrodynamics, and bremsstrahlung {\em per se} is an important phenomenon in many branches of physics.
The exact treatment of bremsstrahlung in a Coulomb potential, as seen in the collision of two charged particles, is a difficult endeavour since it involves the evaluation and integration of hypergeometric functions as part of the positive energy solutions of the respective Schr\"odinger equation. 
While for many applications treating the wave functions as plane waves and solving the problem to first order in perturbation theory is of sufficient accuracy, say for scattering of fast electrons on protons or light ions, there is a significant kinematic regime where such approximation fails. 

Corrections to perturbative results can generally be expected in the non-relativistic regime, {\it i.e.},~when the relative incoming velocity falls below the mutual interaction strength. While there is a vast amount of literature that studies electron-ion bremsstrahlung in this kinematic regime and within a principal astrophysical context \citep{1935MNRAS..96...77M, 1961ApJS....6..167K, 1972ApJ...174..227J, 1975ApJ...199..299K, 1988ApJ...327..477H, vanHoof:2014bha, Chluba:2019ser},
it is perhaps surprising to note that a complete description of electron-electron bremsstrahlung scattering to leading order in the electrons' non-relativistic velocity and exact in their Coulomb interaction is seemingly absent in the literature today.

For the scattering of two particles with different charge-to-mass ratios, the leading contribution to bremsstrahlung is dipole radiation. The classical electrodynamics calculation was given by~\cite{Kramers:1923}. 
\lss{
The first quantum mechanical calculations were done independently by~\cite{Oppenheimer1929} and~\cite{Sugiura1929} in the Born approximation. The full non-relativistic quantum mechanical solution for dipole bremsstrahlung was first found in XYZ form by~\cite{Sommerfeld1931} and in single-differential form by~\cite{Sommerfeld:1935ab}, closing the gap between the Born and the classical limit.
}
A few years later, \cite{Elwert:1939km} found an approximate solution for dipole bremsstrahlung by multiplying the Born cross section by the ratio of probabilities for finding the two initial and final particles, respectively, at the same position. This factor is always greater than unity for attractive and less than unity for repulsive interactions. Thus, the Elwert approximation extends the range of validity of the Born approximation in an easy way. There is, however, still an entire kinematic range where the Elwert approximation fails and the full result must be used. In quest of a generally valid formula for soft photon emission, \cite{Weinberg:2019mai} most recently revisited the process of dipole radiation in an application of his soft photon theorem~\citep{Weinberg:1965nx}.

In case the two scattering particles have the same charge-to-mass ratio, their combined dipole moment vanishes and the leading order contribution to bremsstrahlung is quadrupole radiation.  The first calculation of quadrupole bremsstrahlung for spin-1/2 particles in the Born approximation was done by~\cite{Fedyushin:1952hg}, and later extended to arbitrary spin by~\cite{1981PhRvA..23.2851G}; see also~\citep{1990ApJ...362..284G}. The Elwert prescription was applied to electron-electron bremsstrahlung by~\cite{Maxon1967}, and there is a  body of literature that is concerned with various aspects of the elementary process of electron-electron bremsstrahlung such as the evaulation of Gaunt factors for astrophysical purposes or differential particle distributions as they are relevant in laboratory measurements; we refer the reader to the textbook discussions~\citep{gould2006electromagnetic,Haug:2004gp} and references therein. However, while the exact quantum mechanical formulation of dipole bremsstrahlung has found its entry in advanced textbooks on quantum electrodynamics~\citep{Sommerfeld_Spektrallinien2,akhiezer1953quantum,berestetskii1982quantum},  the corresponding result for quadrupole bremsstrahlung has not yet been presented in a complete manner. 

In this paper, we bring this topic to a closure by calculating the exact differential cross section for quadrupole bremsstrahlung in the non-relativistic approximation and provide its analytical form. Contact is then made to all relevant kinematic regimes, \textit{i.e.}, the Born, classical and soft-photon regimes where our findings can be compared with results from the literature, primarily in connection with the process of electron-electron bremsstrahlung.
For the sake of better generality, we shall consider the collision of
two particles with arbitrary charge-to-mass ratios $Z_1e/m_1$ and
$Z_2e/m_2$ where $e>0$ is the unit of electromagnetic charge. Results are then easily specialized to the case of colliding
electrons $e^-$ and/or positrons $e^+$ of mass $m_{1,2}=m_e$ and charges $Z(e^\pm)=\pm1$, respectively.

The paper is organized as follows: in Sec.~\ref{sec:matr-elem-cross} we introduce the formul\ae\ for the matrix elements and cross sections and in Sec.~\ref{sec:dipole} we briefly review dipole emission in a Coulomb potential. In Sec.~\ref{sec:quadrupole} we then derive the double differential cross sections in photon energy and scattering angle for quadrupole emission. In Sec.~\ref{sec:bornandclass} we relate our results to asymptotic expressions in the Born and classical regime, and in Sec.~\ref{sec:softphoton} we study the regime of soft photon emission. In Sec.~\ref{sec:eff_en_loss} we show numerical results on the effective energy loss, and in Sec.~\ref{sec:gaunt} we propose a definition of a Gaunt factor for electron-electron scattering. We conclude in Sec.~\ref{sec:conclusions} and several appendices provide further details on the calculations.

Throughout the paper we will use rationalized natural units where $\hbar = c  = 1$ and $\alpha=e^2/(4\pi)\simeq 1/137$.

\section{Matrix elements and cross sections}
\label{sec:matr-elem-cross}

The $S$-matrix element for the emission of a photon of energy $\omega$, three-momentum $\vec q$ and helicity $\lambda$ with transverse polarization vector $ {\bm \epsilon}^*_{\vec q,\lambda}$ in Coulomb gauge in the collision of non-relativistic particles with charges $Z_n e$, masses $m_n$, momenta $\vec p_n$ and respective coordinates $\vec x_n$ is given by~\cite{weinberg2015lectures},
\begin{align} \label{eq:s-matrix}
  S_{f\gamma,i} & =  2\pi \, i\,  \delta(E_i - E_f -\omega)
                  \nonumber \\ &  \times
 \frac{ {\bm \epsilon}^*_{\vec q,\lambda}}{\sqrt{2\omega}} \cdot
  \sum_n \frac{Z_n e}{m_n} \langle \Psi_f | e^{-i \vec q \cdot \vec x_n}  \vec p_n | \Psi_i \rangle .
\end{align}
As for our premise of providing a result of quadrupolar photon emission that is exact to all orders in the Coulomb interaction of two colliding particles, we take the inital (final) states $|\Psi_{i(f)}\rangle$ to  be the positive energy solutions of the Hamiltonian
\begin{align} \label{eqn:Hamiltonian}
  H_0 =  \frac{\vec p_1^2}{2m_1} + \frac{\vec p_2^2}{2m_2}+\frac{ Z_1 Z_2\alpha}{|\vec x_1 - \vec x_2|} ,
\end{align}
such that $H_0 | \Psi_{i,f} \rangle = E_{i,f}| \Psi_{i,f} \rangle $.
The motion of the center-of-mass (CM) can  be separated as usual,
$\langle \vec x | \Psi_{i} \rangle = e^{i \vec P\cdot \vec X}
\chi_{\vec p_i}(\vec r)$ where $\vec P = \vec p_1 + \vec p_2$ and
$\vec X = (\mu/m_2) \vec x_1 + (\mu/m_1) \vec x_2$ are CM momentum and coordinate, respectively, and
$\vec r = \vec x_1 - \vec x_2$ and
$\vec p_i = \mu(\vec p_1/m_1 - \vec p_2/m_2)$ are the respective
particles' relative position and initial relative momenta;  $\mu = m_1m_2/(m_1+m_2)$ is the reduced mass. The separation
yields an overall momentum-conserving delta function
$(2\pi)^3\delta^{(3)}(\vec P - \vec p_3 - \vec p_4 - \vec q)$ where
$\vec p_{3,4}$ are the final state momenta with relative momentum
$\vec p_f = \mu(\vec p_3/m_1 - \vec p_4/m_2)$. From the above
matrix element, and in the CM frame $\vec P=0$, we are hence left to
evaluate
\begin{align}
  \label{eq:MatrixEl}
  \vec M_{I} = e \frac{A_I}{\mu} \int d^3\vec r \, \chi_{\vec p_f}^{*}(\vec r) \mathcal{O}_I  (-i\nabla_{\vec r}) \chi_{\vec p_i}(\vec r) .
\end{align}
The operator $ \mathcal{O}_I $ originates from the expansion of the exponentials $e^{-i\vec q \cdot \vec r_n}$ where $\vec r_n = \vec x_n -\vec X$ and the dipole and quadrupole emission cases assemble themselves through the respective leading and next-to-leading order terms,
{
\medmuskip=1mu
\thinmuskip=1mu
\thickmuskip=1mu
\nulldelimiterspace=1pt
\scriptspace=1pt
\begin{align}
  \text{dipole:} & \quad \mathcal{O}_{D} = 1, \quad A_D = \mu\left( \frac{Z_1}{m_1} - \frac{Z_2}{m_2}  \right), \\
   \text{quadrupole:} & \quad \mathcal{O}_{Q} = -i \vec q\cdot \vec r, \quad A_Q =  \mu^2 \left( \frac{Z_1}{m_1^2} + \frac{Z_2}{m_2^2}  \right).
\end{align}
}%
As is evident, if two particles have the same charge-to-mass ratio, $Z_1/m_1=Z_2/m_2$, their mutual dipole moment vanishes and the leading order contribution to bremsstrahlung becomes quadrupole radiation. Hence, for the special cases of scattering of electrons/positrons with their counterparts we have $|A_D^{(e^\pm e^\pm)}|= 0$,   $|A_D^{(e^- e^+)}|= 1$,  $|A_Q^{(e^\pm e^\pm)}|= 1/2$, $|A_Q^{(e^- e^+)}| = 0$.

The wave functions $\chi_{\vec p_{i,f}}$ for the relative motion of the particles that are to be used in (\ref{eq:MatrixEl}) are (linear combinations of) the Coulomb wave functions of appropriate asymptotic behavior~\citep{landau1977quantum}, 
{
\medmuskip=1mu
\thinmuskip=0mu
\thickmuskip=2mu
\nulldelimiterspace=1pt
\scriptspace=0pt
\begin{subequations}
  \label{eq:coulombwaves}
	\begin{align}
		\chi_{\vec p_i}^{(+)}(\vec r) &= \Gamma \left(1-i\nu_i\right)\; e^{\frac{\pi}{2} \nu_i}  e^{i \vec p_i \cdot \vec r} \; {}_1F_1 \!\left(i\nu_i, 1; i\zeta_i^- \right) \,, \\
		\chi_{\vec p_f}^{(-)}(\vec r) &=\Gamma \left(1+i\nu_f\right)\; e^{\frac{\pi}{2} \nu_f}  e^{i \vec p_f \cdot \vec r} \; {}_1F_1 \!\left(-i\nu_f, 1; -i\zeta_f^+ \right) \,,
	\end{align}
\end{subequations}
}%
where ${}_1F_1$ is the  confluent hypergeometric function~\citep{abramowitz1948handbook}.
The magnitudes of the relative three-momenta and position are denoted by  $p_{i,f}\equiv |\vec p_{i,f}|$ and $r\equiv |\vec r|$, respectively, and  $\zeta_{i,f}^\pm=p_{i,f} r \pm \vec p_{i,f} \cdot \vec r$. The relative velocities---small parameters in this problem---are denoted by $v_{i,f}=p_{i,f}/\mu$. Note that we define the Sommerfeld parameters $\nu_{i,f}$ such that they carry the sign of the interaction, 
\begin{align} \label{eqn:nuif}
    \nu_{i,f} \equiv -Z_1Z_2\frac{\alpha}{v_{i,f}} ,
\end{align}
\textit{i.e.},~$\nu_{i,f}>0$ for attractive  interactions ($Z_1 Z_2<0$) and $\nu_{i,f}<0$ for repulsive  interactions ($Z_1 Z_2>0$).
  The wave functions are normalized as $\int d^3\vec r \, \chi_{\vec p}^{*}(\vec r)  \chi_{\vec p'} (\vec r)= (2\pi)^3 \delta^{(3)}(\vec p - \vec p')$ and the asymptotic forms of $\chi^{(+)}$ and $\chi^{(-)}$ at spatial infinity comprise a (Coulomb-distorted) plane wave plus an outgoing and incoming spherical wave, respectively.

The parameters $|\nu_{i,f}|$ can be viewed as the ratio of the magnitude of the potential energy of the colliding particle pair, $|Z_1 Z_2|\alpha /r $, at a separation $r$ that equals the de~Broglie wavelength of the relative motion, $1/(\mu v_{i,f})$, to the respective kinetic energies $\mu v_{i,f}^2/2$. This highlights the role $\nu_{i}$ and $\nu_{f}$ play in delineating the various kinematic regimes, 
\begin{align}
|\nu_{i,f}| & \ll 1 \quad \text{Born regime}, \\ 
|\nu_{i,f} | & \gg 1 \quad \text{classical regime} .
\end{align}
In the Born regime, the electrostatic interaction energy of the two colliding charges before and after the collision remains small compared to the respective kinetic energies. The process then becomes treatable by replacing $\chi_{\vec p_{i,f}}$ with plane waves. In turn, in the opposite limit the particles' motion is quasi-classical and the process becomes describable in terms of formul\ae\ derived from classical electrodynamics; note that $|\nu_f| \geq |\nu_i|$ out of kinematic reasons. Below we shall study those limits and see how the known results follow from the full quantum mechanical formula.

The matrix element $\mathcal{M}$, defined through $S_{f\gamma,i} = (2\pi)^4 \delta(E_i-
E_f-\omega)\delta^{(3)}(\vec P - \vec p_3 - \vec p_4 - \vec q) \mathcal{M} $, when squared, summed over the two  photon polarizations, and averaged over the direction of the outgoing photon is given by
\begin{align} \label{eqn:squared_amplitude}
  \langle | \mathcal{M}_I|^2 \rangle_{\hat q} =  \frac{1}{4\pi} \int d\Omega_{\vec q}  \sum_{\lambda} \left| \frac{ {\bm \epsilon}^*_{\vec q,\lambda}}{\sqrt{2\omega}} \cdot   \vec M_I \right|^2.
\end{align}
The energy-differential cross section for dipole ($I=D$) or quadrupole ($I=Q$) emission is then given by
\begin{align}
\label{eqn:diff_cross_section}
  	\omega \frac{d\sigma_I}{d\omega} &= \frac{\mu^2\omega^3}{4 \pi^3} \sqrt{1- \frac{2\omega}{\mu v_i^2}}
  \int d\cos\theta \; \langle | \mathcal{M}_I|^2 \rangle_{\hat q} \,. 
\end{align}
where $\cos \theta = \vec p_i \cdot \vec p_f/(p_i p_f)$ is the cosine of the CM scattering angle of the colliding particle pair. The boundaries of the $d\cos\theta$ integration are $(-1,1)$ for non-identical particles and $(0,1)$ for identical particles. 

Before proceeding, we comment on the role of spin in the bremsstrahlung process. In the above formulation, all quantities are scalars. The non-relativistic  absence  of  an  interaction  involving  the spin-coordinate  of  the  particles  implies that  the  spin only enters in the counting of spatially symmetric and anti-symmetric states, to be detailed below.  Spin is a relativistic concept and it generally manifests itself in contributions carrying additional factors of $v_{i,f}$ that are hence suppressed. The $S$-matrix element~\eqref{eq:s-matrix} given in  Coulomb gauge arises from the interaction Hamiltonian of charges~$e_n$ with the vector potential~$\vec A$, $H_n = -(e_n/m_n) \vec A \cdot \vec p_n$. This interaction is supplemented by the one arising from the magnetic moment, $H'_n = - \boldsymbol \mu_n \cdot ({\rm curl}\, \vec A)$ where $ |\boldsymbol \mu_n |= e_n/2m_n $ is the intrinsic Dirac moment of a charged spin-1/2 particle. On naive dimensional grounds, ${\rm curl\ } \vec A \sim |\vec q| \vec A  = \omega \vec A$, so that $H'_n/H_n \sim \omega / |\vec p_n| = O(v_{i,f})$, indicating the non-relativistic relative suppression.

In the scattering of identical particles, the leading order process is quadrupole radiation and of higher order in $v_i$ compared to dipole emission. This arises because of a cancellation of the lower order terms among the individual amplitudes in Eq.~\eqref{eq:s-matrix}. A similar cancellation also occurs for the magnetic moment interaction and the relative unimportance of $H'_n$ over $H_n$ is preserved~\citep{1981PhRvA..23.2851G}.
For non-identical particles, dipole emission is the leading order process and radiation induced by $H'_n$ can at best compete with quadrupole emission, but will generally be further suppressed because of a non-divergence in photon energy~\citep{1990ApJ...362..284G}.   
We leave a study of the exact spin-induced emission process in a Coulomb field for dedicated future work, noting that for the primary case of interest, the scattering of identical particles, Eq.~\eqref{eq:s-matrix} yields the dominant result.

\section{Dipole emission}
\label{sec:dipole}

In order to prepare for the quadrupole case and to set some of the notation, in this section we review the calculation of the dipole transition integral.
By defining $\nu_{i,f}$ as we did in Eq.~\eqref{eqn:nuif}, we can treat the case of an attractive Coulomb interaction $Z_1Z_2<0$ and a repulsive Coulomb interaction $Z_1Z_2>0$ simultaneously. 
The wave functions $\chi_{\vec p_i}(\vec r)$ and $\chi_{\vec p_f}(\vec r)$ to be used in (\ref{eq:MatrixEl}) are the ones of (\ref{eq:coulombwaves})~\citep{berestetskii1982quantum},
	\begin{align} \label{eqn:dipole_transition_amp}
		\vec M_D 
		= e \frac{A_D}{\mu} \WFnorm_i \WFnorm_f\lim_{\lambda \rightarrow 0}  & \left[i p_i \nabla_{\vec p_i}  J_1^D(\lambda,\vec k,\vec p_i, \vec p_f) \right. \nonumber \\ &  \left. + \vec p_i  J_2^D(\lambda,\vec k,\vec p_i, \vec p_f)\right] \,,
	\end{align}
with $\WFnorm_{i,f} = e^{-\pi \nu_{i,f}/2} \Gamma \left(1-i\nu_{i,f}\right)$ and using the abbreviation $\Fconf\left(i\zeta_{i(f)} \right)\equiv \!\Fconf\left(i\nu_{i(f)}, 1, i\zeta_{i(f)} \right) $, $ J_1^D$ and $ J_2^D$ are given by
{
\medmuskip=1mu
\thinmuskip=0mu
\thickmuskip=2mu
\nulldelimiterspace=1pt
\scriptspace=0pt
\begin{subequations}
	\begin{align}
		  J_1^D (\lambda,\vec k,\vec p_i, \vec p_f)
		&=\int d^3\vec r \,\frac{e^{i \vec k \cdot \vec r -\lambda r}}{r}\;
		\Fconf(i\zeta_f^+ ) \,
		\Fconf(i\zeta_i^- )\,, \\
		 J_2^D (\lambda,\vec k,\vec p_i, \vec p_f)
		&=\int d^3\vec r\, e^{i \vec k \cdot \vec r-\lambda r}\;
		\Fconf(i\zeta_f^+ ) \,
		\Fconf(i\zeta_i^- )\,.
	\end{align}
\end{subequations}	
}%
The solution to the first integral has been given by~\cite{Nordsieck:1954hg}, which we shall denote by $\mathcal I_N$, \textit{i.e.},~$ J_1^D  = \mathcal I_N$,
{
\medmuskip=0mu
\thinmuskip=0mu
\thickmuskip=0mu
\nulldelimiterspace=2pt
\scriptspace=0pt
	\begin{align}
	\label{eqn:nordsieck_integral}
		\mathcal I_N (\lambda,\vec k,\vec p_i, \vec p_f) &= \frac{2 \pi}{\eta_1} e^{-\pi \nu_i} \left(\frac{\eta_1}{\eta_3}\right)^{i \nu_i} \left(\frac{\eta_3 + \eta_4}{\eta_3}\right)^{-i \nu_f}\nonumber \\
		& \times 
		{}_2F_1\bigg(1-i\nu_i,i\nu_f, 1, \frac{\eta_1 \eta_4 - \eta_2 \eta_3}{\eta_1 (\eta_3+\eta_4)} \bigg) \,,		
	\end{align}
	}%
with $\eta_1 = (k^2+\lambda^2)/2$, $\eta_2 = \vec p_f \cdot \vec k - i \lambda p_f$, $\eta_3 = \vec p_i \cdot \vec k + i \lambda p_i - \eta_1$ and $\eta_4 = p_i p_f + \vec p_i \cdot \vec p_f - \eta_2$. 
The integral $J_2^D$ vanishes due to the orthogonality of the Coulomb wave functions; this is also manifest in the integrated form by noting that $J_2^D = -\partial_\lambda J_1^D$ and $\lim_{\lambda\to 0}\partial_\lambda \mathcal{I}_N =0$. Setting $\vec k = \vec p_i - \vec p_f$ and using the properties of the hypergeometric functions, ${}_2F_1\left( a,b;c;z\right) = (1-z)^{-b} {}_2F_1\left( b,1-a;c;z/(z-1)\right)$ and ${}_2F_1\left( a,b;c;z\right) = {}_2F_1\left( b,a;c;z\right)$
one finds that the dipole transition amplitude is
	\begin{align}
		\vec M_D &=
		\frac{-i 8 \pi e A_D \WFnorm_i \WFnorm_f e^{-\pi \nu_i}}{\mu (p_i-p_f)^2(p_i+p_f)^2}
		\left[\frac{p_i-p_f}{p_i+p_f}\left(1-z\right)\right]^{i\left(\nu_i+\nu_f\right)-1}\nonumber \\
		&\times
		\left[
		i \nu_i p_i  F\left(z\right) \vec k -  (1-z) \left(p_f \vec p_i- p_i \vec p_f\right) F'(z) 
		\right] \,, \notag
	\end{align}
        with $z=-2(p_ip_f-\vec p_i \cdot \vec p_f)/(p_i-p_f)^2$ and where we have introduced the shorthand notation
        \begin{align}
          F(z) &\equiv {}_2F_1(i\nu_f, i\nu_i;1;z)  , \\
          F'(z)& \equiv \frac{\partial}{\partial z}\,  {}_2F_1(i\nu_f, i\nu_i;1;z) ,
        \end{align}
        to be used extensively below. The squared integral is therefore~\citep{berestetskii1982quantum}
{
\medmuskip=1mu
\thinmuskip=1mu
\thickmuskip=1mu
\nulldelimiterspace=1pt
\scriptspace=1pt
	\begin{align} \label{eqn:dipole_squared}
		\mathcal |\vec M_D&|^2 =
		\frac{16 \pi \alpha^3 Z_1^2 Z_2^2 A_D^2 S_i S_f}{(p_i+p_f)^2(p_i-p_f)^4} 
		\bigg[
		\frac{|F\left(z\right)|^2}{1-z}  - \frac{z|F'(z)|^2}{\nu_i \nu_f } \nonumber  \\
		& \!\!\!+i \frac{\nu_i+\nu_f}{2\nu_i \nu_f}\frac{z}{1-z}\Big(F\left(z\right)F'^*(z)-F^*\left(z\right)F'(z)\Big)
		\bigg],
	\end{align}
}%
        where the Sommerfeld factors $S_{i,f}$ characterize the action of the Coulomb field near the origin,
\begin{align} \label{eqn:sommerfeld_enhancement_factor}
S_{i}
&= e^{-2\pi \nu_i} |\chi^{(+)}_{\vec{p}_i}(0)|^2
= \frac{2\pi \nu_{i} }{e^{ 2\pi\nu_{i}}-1 } ,\\
S_{f}
&=  |\chi^{(-)}_{\vec{p}_f}(0)|^2
= \frac{- 2\pi \nu_{f} }{e^{- 2\pi\nu_{f}}-1} .
\end{align}

The differential cross section with respect to the emitted energy $\omega$ is given by effecting the angular integral in \eqref{eqn:dipole_squared}, which can be written as 
$d \cos \theta = (p_i-p_f)^2/(2 p_ip_f) \, dz $.
In the dipole case, one can then use the hypergeometric differential equation to substitute the expression in the square brackets of (\ref{eqn:dipole_squared}) by a total derivative in $z$
to get the differential cross section for an attractive interaction~\citep{berestetskii1982quantum}
	\begin{align} \label{eqn:full_quantum_cs}
		\omega \frac{d\sigma_D}{d\omega} = \frac{16}{3} \frac{\alpha^3 Z_1^2 Z_2^2 A_D^2}{\mu^2 v_i^2} 
		S_i S_f \;
		\frac{1}{\nu_i \nu_f} \frac{\xi}{4} \frac{d}{d\xi}\left|F(\xi)\right|^2 \,,
	\end{align}
with $\xi=\min(z)= -4p_ip_f/\left(p_i-p_f\right)^2$.  Note that $\nu_{i,f}$, arguments in the Sommerfeld factors $S_{i,f}$ and parameters in the hypergeometric function $F$, are positive for attractive interactions and negative for repulsive interactions. Equation~\eqref{eqn:full_quantum_cs} is therefore  applicable to both cases.

\section{Quadrupole emission}
\label{sec:quadrupole}

If two particles have the same charge-to-mass ratio, their combined dipole moment vanishes and the leading order contribution to bremsstrahlung is quadrupole radiation. The  quantum mechanical calculation for the collision of two non-relativistic electrons to first order in perturbation theory was presented by~\cite{Fedyushin:1952hg}; for  textbook discussions see~\cite{berestetskii1982quantum,akhiezer1953quantum}.
We now proceed to the calculation that is complete to all orders in the Coulomb interaction of the scattering particles. We additionally recover the previous results by~\cite{Fedyushin:1952hg} in a dedicated Born-level quantum field theory calculation that is presented in App.~\ref{sec:mat_elements}.

We start this section by considering the scattering of a
distinguishable particle pair. In this case, $A_D\neq 0$, dipole radiation dominates (unless $A_D\ll A_Q$ because of some accidental cancellation), and quadrupole radiation is the first correction in a velocity expansion of the cross section. In a second part, we then consider the scattering of identical particles, such as two electrons, for which $A_D=0$ and quadrupole radiation will become the leading process.  

The calculations below will be framed in terms of a quadrupole tensor $\mathbfcal Q = (Q^{ab})$ that is defined through the
expression~(\ref{eq:MatrixEl}) for the matrix element $\vec{M}_Q = (M_Q^b) $, 
\begin{align} 
   M_Q^b = - \frac{A_Q}{\mu}\, q^a  Q^{ab}    . 
\end{align}
The Cartesian components of $\mathbfcal Q$ are then given by the  tensor\footnote{Throughout this paper, Cartesian coordinates are given by upper indices $a,b,\dots$ to discern them from the lower indices $i$ and $f$ labeling in and out states; the Einstein summation convention is used for upper indices.} 
\begin{align}
  \label{eq:QmnDef}
 Q^{ab} = e\int d^3\vec r \, \chi_{\vec p_f}^{*}(\vec  r)  r^a \partial^b \chi_{\vec p_i}(\vec r),
\end{align}
where $\partial^b$ is the derivative with respect to the $b$-th
Cartesian component of $\vec r$, \textit{i.e.}~$r^b$. The tensor is symmetric in its indices, $Q^{ab}= Q^{ba}$, which is related to the fact that the two-particle system (without explicit magnetic moment interaction)
has vanishing magnetic dipole moment, see \textit{e.g.}~\cite{landau1975classical}.

The squared matrix element \eqref{eqn:squared_amplitude} can be then written in terms of the quadrupole tensor as
\begin{align}
    \langle | \mathcal{M}_Q|^2 \rangle_{\hat q} &=
    \frac{\omega A_Q^2}{2\mu^2 }  \Big\langle \sum_{\lambda}\thinhat q^a \thinhat q^c (\epsilon_{\vec q,\lambda}^*)^{b} \epsilon_{\vec q,\lambda}^d \Big\rangle_{\hat q} (Q^*)^{ab} Q^{cd} \nonumber \\
	&=	\frac{\omega A_Q^2}{30\mu^2} 
		 \left[3 | Q^{ab}|^2 - | Q^{aa}|^2\right]
    \,, \label{eqn:transition_amp_vector}
\end{align}
with $| Q^{ab}|^2\equiv (Q^*)^{ab} Q^{ba} = \Tr(\mathbfcal{Q}^\dag \mathbfcal{Q})$ and $| Q^{aa}|^2 \equiv (Q^*)^{aa} Q^{bb} = | \Tr( \mathbfcal{Q}) |^2$. In the second equality
 we have used for the angular average
\begin{align}
   \Big\langle \sum_{\lambda}\thinhat q^a \thinhat q^c \epsilon^{*b} \epsilon^d \Big\rangle_{\hat q} = &  \frac{3}{30}\left(\delta^{ac}\delta^{bd}+ \delta^{bc}\delta^{ad} - \frac{2}{3} \delta^{ab}\delta^{cd} \right) \nonumber \\
  & + \frac{5}{30}\left( \delta^{ac}\delta^{bd} - \delta^{bc}\delta^{ad}  \right). 
\end{align}
At this point the quadrupolar E2 nature of the emission becomes manifest: because of the symmetry in $Q^{ab}$, the contraction with the second bracket vanishes and only the symmetric combination in $a$ and $b$ and in $c$ and $d$ survives. The term corresponding to magnetic dipole emission M1 is therefore not present.

An important aspect in the calculation of $\mathbfcal Q$ is that for identical spin-1/2 particles, the overall wave function is anti-symmetric under particle exchange. 
 If the particles are in the spin singlet ($\singlet$) state, the  wave-function is spatially symmetric and vice versa for the three spin triplet ($\triplet$) states. The wave functions $\chi_{\vec p_{i,f}}$ to be used are then,
\begin{subequations}
\label{eq:symmwf}
  \begin{align}\label{eq:symm_wf}
    \chi^{\singlet}_{\vec p_i} (\vec r) & =  \frac{1}{\sqrt{2}}
\left(\chi_{\vec p_i}^{(+)}(\vec r) + \chi_{\vec p_i}^{(+)}(-\vec r) \right)\,, \\
   \chi^{\triplet}_{\vec p_i} (\vec r) & =  \frac{1}{\sqrt{2}}
\left(\chi_{\vec p_i}^{(+)}(\vec r) - \chi_{\vec p_i}^{(+)}(-\vec r) \right)\,, 
  \end{align}
\end{subequations}
with analogous linear combinations  $\chi^{\singlet}_{\vec p_f} (\vec r)$ and $\chi^{\triplet}_{\vec p_f} (\vec r)$  made from $\chi_{\vec p_f}^{(-)}$. 
For identical spin-0 particles, the wave function is symmetric under particle exchange and the wave function \eqref{eq:symm_wf} has to be used.
For the scattering of identical particles, there are hence more overlap integrals of hypergeometric functions to consider.

\subsection{Scattering of non-identical particles} \label{sec:quad_distinguishable_particles}

If the scattered particles are distinguishable, the wave functions $\chi_{\vec p_i}(\vec r)$ and $\chi_{\vec p_f}(\vec r)$ to be used in (\ref{eq:QmnDef}) are the ones of (\ref{eq:coulombwaves}).
 Following the notation for the dipole case, we replace the scalar functions $ J_{1,2}^D$ with vector-valued functions $\mathbf J^Q_{1,2}$, given by
\begin{subequations}
	\begin{align}
		\mathbf J_1^Q (\lambda,\vec k,\vec p_i, \vec p_f) 
		&=\int d^3\vec r\, \frac{\vec r}{r} e^{i \vec k \cdot \vec r-\lambda r} 
		\Fconf( i\zeta_f^+ ) 
		\Fconf(i \zeta_i^- ) \nonumber \\
		&=  -i  \nabla_{\vec k} \mathcal I_N(\lambda,\vec k,\vec p_i, \vec p_f), \label{eqn:J1Q}
           \\
		\mathbf J_2^Q (\lambda,\vec k,\vec p_i, \vec p_f)
		&= \!\int\!\! d^3\vec r \,\, \vec r \, e^{i \vec k \cdot \vec r-\lambda r}
		\Fconf( i\zeta_f^+ ) 
		\Fconf(i \zeta_i^- )\nonumber \\
		&=i \frac{\partial}{\partial\lambda} \nabla_{\vec k}  \mathcal I_N(\lambda,\vec k,\vec p_i, \vec p_f). \label{eqn:J2Q}
	\end{align}
\end{subequations}
We hence make the important observation, that the integrals are expressible as derivatives of the Nordsieck integral $\mathcal I_N$ given by Eq.~\eqref{eqn:nordsieck_integral}. Because of this, we are in the position of assembling an analytical solution. Also note that unlike in the dipole case, $\mathbf J_2^Q$ does not identically vanish. From the definitions \eqref{eqn:J1Q} and \eqref{eqn:J2Q}, it follows that the quadrupole tensor $\mathbfcal Q$ with Cartesian components $Q^{ab}$ can be written as
{
\medmuskip=0mu
\thinmuskip=0mu
\thickmuskip=0mu
\nulldelimiterspace=2pt
\scriptspace=1pt
\begin{align}\label{QinNordsieck}
	 Q^{ab}
		&=e\,  \WFnorm_i \WFnorm_f \lim_{\lambda \rightarrow 0} \frac{\partial}{\partial k^a} \left[i p_i \frac{\partial}{\partial p_i^b}  -  p_i^b \frac{\partial}{\partial \lambda}\right] \mathcal I_N(\lambda,\vec k,\vec p_i, \vec p_f) \,,
\end{align}
}%
where $\vec p_i$, $\vec p_f$, $\vec k$ are treated as separate variables in the partial derivatives of $\mathcal{I}_N$. After taking the derivatives, we set $\vec k = \vec p_i - \vec p_f$. The elements of the quadrupole tensor can then be split into the following symmetric combinations,
{
\medmuskip=0mu
\thinmuskip=0mu
\thickmuskip=0mu
\nulldelimiterspace=2pt
\scriptspace=0pt
	\begin{align} \label{eqn:quad_tensor}
		Q^{ab} 
		&=
		f_0
		\bigg[
		\left(f_1 \thinhat p_i^a \thinhat p_i^b + f_2 \thinhat p_f^a \thinhat p_f^b + f_3 (\thinhat p_i^a \thinhat p_f^b + \thinhat p_i^b \thinhat p_f^a)\right) F(z)  \nonumber \\
		&\!\!\!\!+
		\left(f_4 \thinhat p_i^a \thinhat p_i^b + f_5 \thinhat p_f^a \thinhat p_f^b + f_6 (\thinhat p_i^a \thinhat p_f^b + \thinhat p_i^b \thinhat p_f^a)\right) F'(z)
		\bigg] ,
	\end{align}
}\\
where the various $\thinhat p_{i,f}^a$ are the Cartesian components of the unit vectors  $\hat p_{i,f} = \vec p_{i,f}/p_{i,f}$. The prefactors $f_0$ to $f_6$ are
\begin{subequations} \label{eqn:quad_tensor_coefficients}
	\begin{align}
		f_0 &=
		\frac{1}{p_i^3} \frac{e\, \WFnorm_i \WFnorm_f\; 16 \pi   e^{-\pi \nu_i}}{(1-y)^3(1+y)^4}
		\left[\frac{1-y}{1+y}\left(1-z\right)\right]^{i\left(\nu_i+\nu_f\right)-2} 
		\!\!\!\!\!\!\!\!\!\!\!\!\!\!\!\!\!\!\!\!\!\!\!\!, \\
		f_1 &=
		\nu_i \Big[-2+(1-y)z\Big]  \nonumber\\
		& \quad +i\frac{\nu_i^2}{z y}\Big[2y^2+2y(2-y)z + (1-y)^2 z^2\Big] \,,
\\
		f_2 &=
		\nu_i \Big[-2y^3-y^2(1-y)z\Big]   \nonumber\\
		&  \quad+i  \frac{\nu_i^2}{z y}\Big[2y^2+2y^2(2y-1)z + y^2(1-y)^2 z^2\Big] \,,
		\!\!\!\!\!\!\!\!
		\\
		f_3 &=
		\nu_i \Big[y(1+y)\Big]     +i \frac{\nu_i^2}{z y} \Big[-2y^2-y(1+y^2)z \Big] \,,
		\\
		f_4 &= 
		\nu_i(1-z)\Big[-y(3-y)-(1-y)^2z\Big]   \nonumber\\
		&  \quad+i \frac{(1-z)^2}{z} \Big[2y^2-y(1-y)z \Big] \,,
		\\
		f_5 &= 
		\nu_i(1-z)\Big[y(1-3y)-y(1-y)^2z\Big]  \nonumber\\
		&  \quad +i \frac{(1-z)^2}{z}  \Big[2y^2+y^2(1-y)z\Big] \,,
		\\
		f_6 &=
		\nu_i(1-z)\Big[y(1+y)\Big]    - i \frac{(1-z)^2}{z} 2 y^2 \,.
	\end{align}
      \end{subequations}
      with $y=p_f/p_i$ and the definitions of $z$ as in Sec.~\ref{sec:dipole} above.

The elements of the quadrupole tensor~\eqref{eqn:quad_tensor_coefficients} for a Coulomb field have previously been reported in \cite{Galstov1976}. The results appear to have gone unnoticed in the Western literature, see \textit{e.g.}~\cite{gould2006electromagnetic} claiming the absence of such calculation (p.\,236). We disagree---by individually different amounts---with all real coefficients of $F$ and two of the three imaginary coefficients of $F'$ listed in~Eq.~(17) of~\cite{Galstov1976}. A numerical  study of the ensuing cross section shows that it generally yields wrong results and asymptotes to the correct Born and classical limits only in a limited region of parameter space. The authors start from a different ansatz---Eqs.~(1) and~(2) by~\cite{Galtsov1974}---to establish the elements of the quadrupole tensor from the Nordsieck integral. Using their starting point, we recover our coefficients~\eqref{eqn:quad_tensor_coefficients} for the quadrupole tensor. 
 From this we conclude that the result of~\cite{Galstov1976} contains errors 
 (English and Russian version are consistent in their critical equation).
 Despite those discrepancies, we agree in the deduced asymptotic expression for the classical limit for $\kappa\gg 1$, Eq.~\eqref{sigma_quad_large_nu_att} in this work.

      At this point, we can calculate $|Q^{ab}|^2 = \Tr(\mathbfcal{Q}^{\dag}\mathbfcal{Q})$ and $| Q^{aa}|^2 = |\Tr(\mathbfcal{Q})|^2$ with $\left(\hat p_i\cdot {\hat p_f}\right)=\cos \theta = 1+z(1-y)^2/(2y)$, both of which can be written in the form%
	\begin{align} \label{eqn:Q}
          | Q^{\alpha\beta}|^2
		&= G_0
		\bigg[
		\left(A_{\thinhat\alpha\thinhat\beta}+C_{\thinhat\alpha\thinhat\beta} \frac{d}{dz}\right) \left|F(z)\right|^2
		+B_{\thinhat\alpha\thinhat\beta} \left|F'(z)\right|^2 \nonumber \\
		&+i D_{\thinhat\alpha\thinhat\beta}\left(F\left(z\right)F'^*(z)-F^*\left(z\right)F'(z)\right)
		\bigg] \,.
	\end{align}
        The hats on $\thinhat\alpha\thinhat\beta = \thinhat a\thinhat a$ or $\thinhat a\thinhat b$ are labels of the coefficients $A$ to $D$ and not Cartesian components that would need to be summed over. 
The prefactor reads
\begin{align} \label{eqn:G_0}
	G_0 &= \frac{64 \pi \alpha^3 Z_1^2 Z_2^2 S_i S_f}{\mu^6 v_i^8 \; (1+y)^4 (1-y)^6} \,.
\end{align}
The  coefficients $A$ to $D$ for $|Q^{aa}|^2$ are given by
{
\medmuskip=1mu
\thinmuskip=0mu
\thickmuskip=2mu
\nulldelimiterspace=2pt
\scriptspace=1pt
\begin{subequations} \label{eqn:Qnn}
	\begin{align}
		A_{\thinhat a\thinhat a} &=
		\frac{4  (y+1)^2}{ (z-1)^2 }, \;\; \;\; \;\;
		B_{\thinhat a\thinhat a} = 
		\frac{9 y^2}{\nu_i^2 } , \;\; \;\; \;\;
		D_{\thinhat a\thinhat a} = 
		\frac{6  y (y+1)}{  \nu_i (z-1)} \,,
	\end{align}
\end{subequations}
}%
and $C_{\thinhat a\thinhat a}=0$. For $|\mathcal Q^{ab}|^2$ we find
{
\medmuskip=1mu
\thinmuskip=0mu
\thickmuskip=2mu
\nulldelimiterspace=2pt
\scriptspace=1pt
\begin{subequations}\label{eqn:Qmn}
	\begin{align}
		A_{\thinhat a\thinhat b} &=
		\frac{\nu_i^2}{2  (z-1)^2} 
		\Bigg\{
		\left[ \frac{(y-1)^4}{y^2}- \frac{ (y-1)^2}{\nu_i^2}\right] z^2 \nonumber\\
		& +
		\frac{4}{y}\left[2 (1-y)^2 - \frac{y^2}{\nu_i^2}\right] z
		+
		8  \left[2 +\frac{(y+1)^2}{\nu_i^2}\right]
		\Bigg\} ,
	\\
		B_{\thinhat a\thinhat b} &= 
		\frac{1}{2y} 
		\Bigg\{
		\frac{ z}{(z-1)} \left[(y-1)^2 z+4 y\right]^2 \nonumber\\
		&+
		\frac{y^2}{\nu_i^2}\left[10 y - 2 \left(1+y^2\right) z - (1-y)^2 z^2\right]
		\Bigg\} , 
	\\
		C_{\thinhat a\thinhat b} &= 
		\frac{ (1-y)^4}{2y (z-1)^2}  \left[ z+ \frac{4 y}{(1-y)^2}\right] 
		 \left[ z^2+\frac{y (1+3z)}{(1-y)^2} \right] ,
	\\
		D_{\thinhat a\thinhat b} &= 
		-\frac{ (1+y)}{4 \nu_i y^2(1-z)^2}  
		\Bigg\{
		\left[ y^2-\nu_i^2 (y-1)^2\right] \nonumber\\
		&\times (y-1)^2 z^3 
		+
		\left[4 y^3-8 \nu_i^2 (y-1)^2 y\right] z^2 \nonumber\\
		&-
		4 y^2 \left[4 \nu_i^2+(y+4) y+1\right] z
		+
		8y^3
		\Bigg\} . 
	\end{align}
\end{subequations}
}%

The double differential cross section in the incoming particles' scattering angle $\theta$ and the emitted energy $\omega$ can then be obtained by plugging \eqref{eqn:Q} into \eqref{eqn:diff_cross_section} using the respective expressions for the quadrupole transition amplitude given by \eqref{eqn:transition_amp_vector},
\begin{widetext}
\begin{align} \label{eqn:dist_cross_section}
     \left. \frac{\omega\, d\sigma_Q}{d\omega\, d\cos\theta} \right|_{\text{non-id.}}&= 
		\frac{8}{15}\frac{\alpha^3 Z_1^2 Z_2^2 A_Q^2}{\mu^2} \frac{y}{ (1-y)^2} S_i S_f
		\bigg[
		\left(A+C \frac{d}{dz}\right) \left|F(z)\right|^2 
		+B \left|F'(z)\right|^2 + 2 D\, \Im\left[F^*\left(z\right)F'(z)\right]
		\bigg] ,
\end{align}
\end{widetext}
where $A=3A_{\thinhat a\thinhat b}-A_{\thinhat a\thinhat a}$ and the
same for $B$, $C$ and $D$.
Again, the expression for the cross section is valid for attractive and repulsive interactions with $\nu_{i}$ changing sign going from the former to the latter.  Equation~(\ref{eqn:dist_cross_section}) is our first main result.

\subsection{Scattering of identical particles}

To calculate the scattering of identical particles---such as in electron-electron scattering---we have to use the spatially (anti-)symmetrized wave functions (\ref{eq:symmwf}). In the following,
we treat the symmetric and antisymmetric case separately, 
\begin{subequations}
	\begin{align}
		Q_{\singlet}^{ab} &= e
		\int d^3\vec r \; (\chi_{\vec p_f}^\singlet )^* \;   r^a \partial^b \; \chi_{\vec p_i}^\singlet \,,\\
		Q_{\triplet}^{ab} &= e
		\int d^3\vec r \; (\chi_{\vec p_f}^\triplet )^* \;   r^a \partial^b \; \chi_{\vec p_i}^\triplet\,.
	\end{align}
\end{subequations}
The squared amplitude will then be given by the singlet combination $| Q^{\alpha\beta}_{\singlet}|^2$ for spin-0 and by
\begin{align}
| Q^{\alpha\beta}|^2 = \frac{1}{4}\big(| Q^{\alpha\beta}_{\singlet}|^2 + 3 | Q^{\alpha\beta}_{\triplet}|^2 \big)  \qquad \alpha \beta =  a a , \, ab, 
\end{align}
for spin-1/2 particles, where the factor $1/4$ is from the spin average of the four possible (singlet and triplet) spin configurations of the colliding unpolarized spin-1/2 particles. 
The four relevant integrals are 
{
\medmuskip=1mu
\thinmuskip=0mu
\thickmuskip=2mu
\nulldelimiterspace=1pt
\scriptspace=0pt
\begin{subequations}
	\begin{align}
		Q_1^{ab} &=
		e\int d^3 \vec r \; \left[\chi_{\vec p_f}^{(-)}(\vec r)\right]^* r^a \partial^b \; \chi_{\vec p_i}^{(+)}(\vec r) \nonumber\\
		&= 
		e\,\WFnorm_i \WFnorm_f \frac{\partial}{\partial k^a} \left[i p_i \frac{\partial}{\partial p_i^b}  -  p_i^b \frac{\partial}{\partial \lambda}\right] \mathcal I_N (\lambda,\vec k,\vec p_i,\vec p_f) \,,
		\\
		Q_2^{ab} &=
		e\int d^3 \vec r \; \left[\chi_{\vec p_f}^{(-)}(-\vec r)\right]^* r^a \partial^b \; \chi_{\vec p_i}^{(+)}(\vec r) \nonumber\\
		&= 
		e\,\WFnorm_i \WFnorm_f \frac{\partial}{\partial l^a} \left[i p_i \frac{\partial}{\partial p_i^b}  -  p_i^b \frac{\partial}{\partial \lambda}\right] \mathcal I_N (\lambda,\vec l,\vec p_i,-\vec p_f) \,,
		\\
		Q_3^{ab} &=
		e\int d^3 \vec r \; \left[\chi_{\vec p_f}^{(-)}(-\vec r)\right]^* r^a \partial^b \; \chi_{\vec p_i}^{(+)}(-\vec r) =Q_1^{ab} \,,
		\\
		Q_4^{ab} &=
		e\int d^3 \vec r \; \left[\chi_{\vec p_f}^{(-)}(\vec r)\right]^* r^a \partial^b \; \chi_{\vec p_i}^{(+)}(-\vec r) =Q_2^{ab} \,,
	\end{align}
\end{subequations}
}%
with $\mathbfcal Q_{\singlet} = (\mathbfcal Q_1 + \mathbfcal Q_2 + \mathbfcal Q_3 + \mathbfcal Q_4)/2$ and $\mathbfcal Q_{\triplet} = (\mathbfcal Q_1 - \mathbfcal Q_2 + \mathbfcal Q_3 - \mathbfcal Q_4)/2$,
where $\vec k = \vec p_i-\vec p_f$ (as before) and $\vec l \equiv \vec p_i+\vec p_f$ are treated as independent from $\vec p_i$ and $\vec p_f$ until after taking the derivative; furthermore  $\lim_{\lambda\to 0}\mathcal I_N (\lambda,-\vec k,-\vec p_i,-\vec p_f)=\lim_{\lambda\to 0}\mathcal I_N (\lambda,\vec k,\vec p_i,\vec p_f)$   holds. We can see that $\mathbfcal Q_1 = \mathbfcal Q_3$ and $\mathbfcal Q_2 = \mathbfcal Q_4$ which gives
    \begin{align} \label{eqn:quad_tensor_identical_particles}
		| Q^{\alpha \beta}|^2 &= | Q_1^{\alpha \beta}|^2 + | Q_2^{\alpha \beta}|^2 + \frac{(-1)^{2s}}{2s+1} |I^{\alpha \beta}|^2 \,,
	\end{align}
where $\alpha \beta =  a b$ or $ a a$  as before and $s$ is the spin of the colliding particles.
Here, $| Q_1^{\alpha \beta}|^2$ is nothing else than the squared quadrupole tensor \eqref{eqn:Q} of Sec.~\ref{sec:quad_distinguishable_particles}. It can be easily shown that $|Q_2^{\alpha \beta}|^2$ can be obtained from $| Q_1^{\alpha \beta}|^2$ by replacing $\cos \theta \to -\cos \theta$, which is identical to the transformation $z \to \tilde z= -z-4y/(1-y)^2$. For the interference term in \eqref{eqn:quad_tensor_identical_particles}, we can use the expression \eqref{eqn:quad_tensor} for the quadrupole tensors $\mathbfcal Q_1$ and $\mathbfcal Q_2$ with $\mathbfcal Q_1 \equiv \mathbfcal Q_1(z)$ and $\mathbfcal Q_2 \equiv \mathbfcal Q_2(\tilde z)$. The result for the interference term $|I^{\alpha \beta}|^2$ is rather lengthy  which we therefore relegate to App.~\ref{sec:coefficients}.
In the limit $\nu_i \to 0$, the expressions obtained by plugging \eqref{eqn:Qnn_id} and \eqref{eqn:Qmn_id} into the three terms in \eqref{eqn:quad_tensor_identical_particles} agree with the expressions obtained from the t-channel, u-channel, and interference terms of the tree level diagrams in a Born-level calculation. 
The double differential cross section for two identical particles with mass $m_{1,2}=m$ and charge $Z_{1,2}=Z$ so that \lss{$A_Q = Z/2$} reads,
\begin{widetext}
{\medmuskip=0mu
\thinmuskip=1mu
\thickmuskip=0mu
\nulldelimiterspace=1pt
\scriptspace=0pt
\begin{align}
    \left. \frac{\omega\,  d\sigma_Q}{d\omega d\cos\theta}\right|_{\text{ident.}} 
    &= 
		\frac{8}{15}\frac{\alpha^3 Z^6}{m^2} \frac{y}{ (1-y)^2} S_i S_f
		\Bigg[
		\left(A+C \frac{d}{dz}\right) \left|F(z)\right|^2 
		+B \left|F'(z)\right|^2 + 2D \, \Im\left[F^*\left(z\right)F'(z)\right]
		+
		\left(\tilde A+ \tilde C \frac{d}{d\tilde z}\right) \left|F(\tilde z)\right|^2 \nonumber\\
		& \!\!\!\!\!\!\!\!\!\!\!\!\!\!\!\!\!\!\!\!\!\!\!\!\!\!\!\!\!\!\!\!\!\!\!\!\!\!\!\!\!\!
		\!\!\!\!\!\!\!\!\!\!\!\!\!\!\!\!\!\!\!\!\!\!\!\!\!\!\!\!\!\!\!\!\!\!\!\!\!\!\!\!\!\!
		+ \tilde B \left|F'(\tilde z)\right|^2
		+ 2 \tilde D \, \Im\left[F^*\left(\tilde z\right)F'(\tilde z)\right]
		 + 2\frac{(-1)^{2s}}{2s+1} 
		\bigg\{
		A^+  \Re\left[F^*(z)F(\tilde z)\right]
		+ 
		A^-  \Im\left[F^*(z)F(\tilde z)\right] 
		+
		B^+\Re\left[F'^*(z)F'(\tilde z)\right]
		+ \nonumber\\
		& \!\!\!\!\!\!\!\!\!\!\!\!\!\!\!\!\!\!\!\!\!\!\!\!\!\!\!\!\!\!\!\!\!\!\!\!\!\!\!\!\!\!
		\!\!\!\!\!\!\!\!\!\!\!\!\!\!\!\!\!\!\!\!\!\!\!\!\!\!\!\!\!\!\!\!\!\!\!\!\!\!\!\!\!\!
		B^- \Im\left[F'^*(z)F'(\tilde z)\right] 
		+
		C^+\Re\left[F^*(z)F'(\tilde z)\right] 
		+ 
		C^-\Im\left[F^*(z)F'(\tilde z)\right] 
		+
		D^+\Re\left[F'^*(z)F(\tilde z)\right]
		+ 
		D^-\Im\left[F'^*(z)F(\tilde z)\right]
		\bigg\}
		\Bigg] \,.
		\label{eq:quad_cs_full_id}
\end{align}}%
\end{widetext}
where, as in Eq.~\eqref{eqn:dist_cross_section},
$A=3A_{\thinhat a\thinhat b}-A_{\thinhat a\thinhat a}$, 
$A^\pm=3A^\pm_{\thinhat a\thinhat b}-A^\pm_{\thinhat a\thinhat a}$, 
$\tilde A = A(z\to \tilde z)$
and the same
for the other coefficients $B,\ C,\ D$ and $\tilde B,\ \tilde C,\ \tilde D$, as well as $B^\pm,\ C^\pm,\ D^\pm$.
This is our second main result, and it is the exact expression for bremsstrahlung emission in the scattering of a non-relativistic electron pair. Note that in the scattering of identical particles, the latter only share repulsive Coulomb interactions.

\begin{figure*}[tb]
		\includegraphics[width=0.95\textwidth]{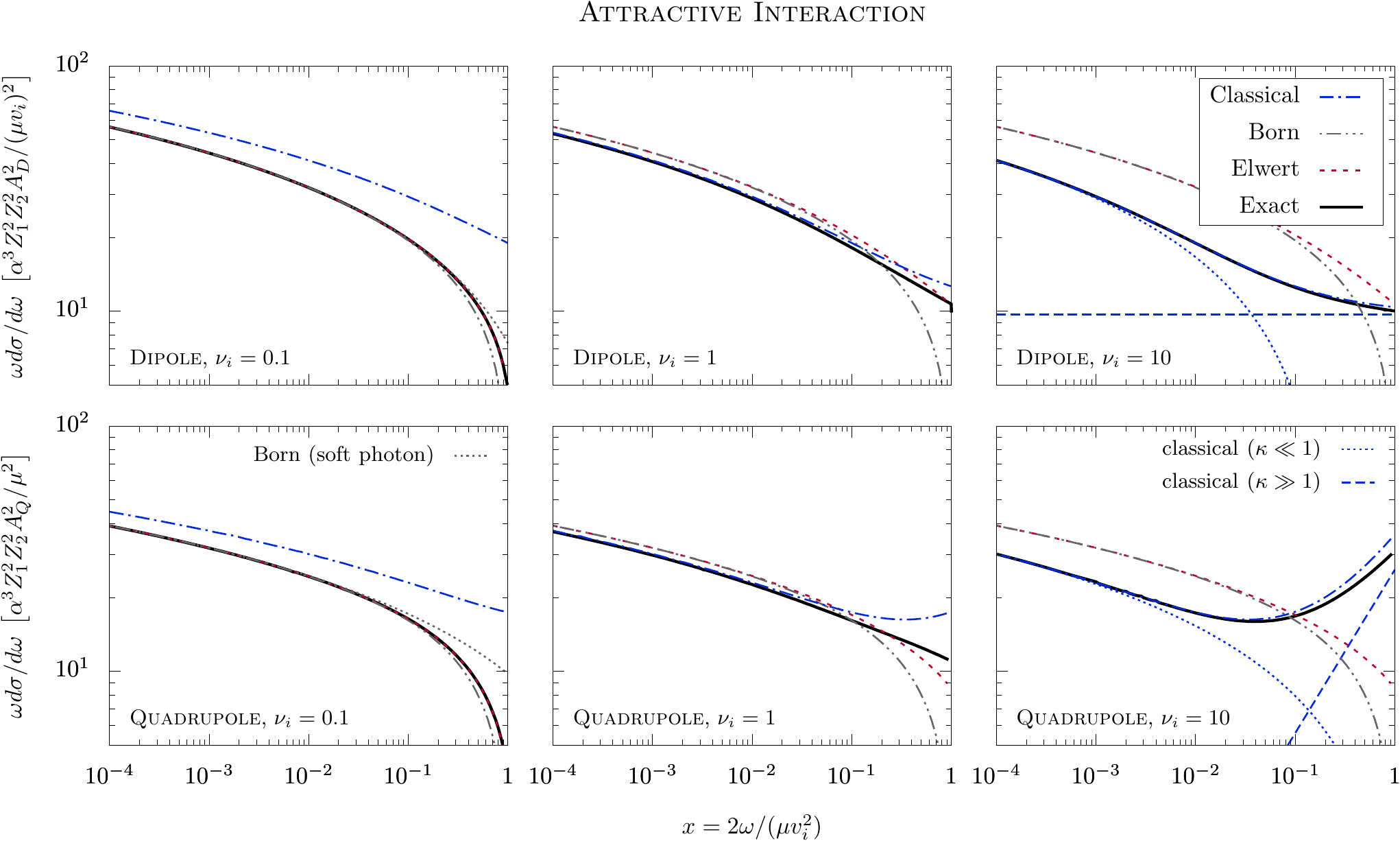}
		\caption{Differential cross section in $\omega$ 
		for attractive interactions $Z_1Z_2<0$ for dipole (top) and quadrupole (bottom) emission for three different values of $\nu_i = \{0.1,1,10\}$ (increasing from left to right). 
		The exact result [Eq.~\eqref{eqn:full_quantum_cs} for dipole emission and angle-integrated Eq.~\eqref{eqn:dist_cross_section} for quadrupole emission] is shown in solid black. The Born limits~\eqref{eqn:cs_vector_pa} and \eqref{eqn:vector_ferm_pp_cs} are given by the dash double-dotted gray lines and the rigorous classical limit [Eq.~\eqref{eqn:classical_cs_full} for dipole and Eq.~\eqref{eqn:eqn:classical_quad_cs} integrated over $\epsilon$ for quadrupole emission] is shown by the dash-dotted blue lines. The classical asymptotic expressions \eqref{sigma_dipole_large_nu_att} and \eqref{sigma_quad_large_nu_att} are shown by the long-dashed blue thin lines and the classical soft photon limit \eqref{sigma_dipole_large_nu_att_soft} and \eqref{sigma_quad_large_nu_att_soft} are given by the thin dotted blue lines in the right panels. In the left panels, the Born soft-photon limits \eqref{eqn:cs_vector_pa_soft} and \eqref{eqn:vector_ferm_pp_cs_soft} are shown by thin dotted gray lines. The Elwert approximation is shown in dashed red. \label{fig:sigma_diff}}
\end{figure*}

\begin{figure*}[tb]
		\includegraphics[width=0.95\textwidth]{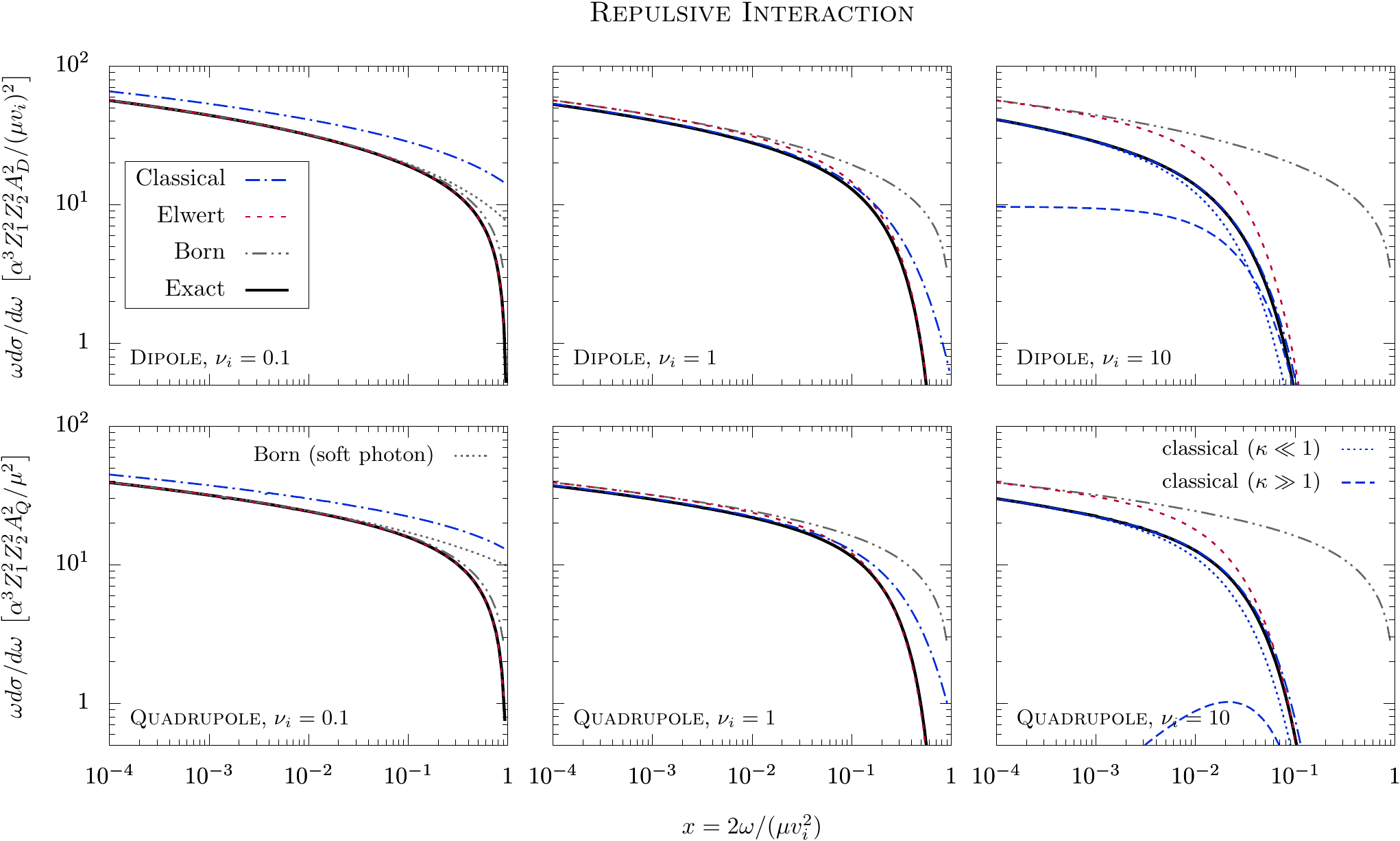}
		\caption{Same as Fig.~\ref{fig:sigma_diff} but for repulsive interactions.  \label{fig:sigma_diff_repulsive}}
\end{figure*}

\section{Born and Quasi-classical Limits}
\label{sec:bornandclass}

The expressions presented in the previous section are valid for bremsstrahlung in the presence of Coulomb interactions for general values of the Sommerfeld parameters $\nu_{i,f}$. Hence, they contain both, the Born and the quasi-classical limit.
In the following, we shall discuss how the limiting expressions, as they are often found in the literature, appear from the exact results above; dedicated Born and classical calculations of the cross sections  are additionally presented in the appendices. 

\subsection{Born Limit} \label{sec:born_limit}

The Born regime is characterized by $|\nu_{i,f}|\ll 1$. In this limit, the hypergeometric function and its derivative can be written in closed form as~\citep{abramowitz1948handbook}
\begin{subequations}\label{eq:BornLimitFs}
\begin{align}
	F(z) &\approx 1 \,, \quad
	 F'(z)  \approx \frac{\nu_i^2}{y} \frac{\ln(1-z)}{z} \quad (|\nu_i|\ll 1) .
\end{align}
\end{subequations}
In the same limit, the Sommerfeld factors go to unity, $S_{i,f}\to 1$. It is then straight-forward to obtain the Born expressions, which we shall record in the following.

For dipole emission, and denoting by $x \equiv 2\omega/\mu v_i^2$ the fraction of kinetic CM energy that is carried away by the emitted photon, the energy differential cross section reads
\begin{align} \label{eqn:cs_vector_pa}
\omega \left. \frac{d\sigma_D}{d\omega} \right|_{\text{Born}}
&= \frac{16}{3} \frac{\alpha^{3} Z_1^2 Z_2^2 A_D^2}{\mu^2 v_i^2}
 \ln \left(\frac{1+ \sqrt{1-x}}{1- \sqrt{1-x}}\right) .
\end{align}

For quadrupole emission, the  corresponding cross section for non-identical particles is obtained from (\ref{eqn:dist_cross_section}) by taking the limits~(\ref{eq:BornLimitFs}) and by integrating over $\cos\theta$,
	\begin{align} \label{eqn:vector_ferm_pp_cs}
		\omega \left.\frac{d\sigma_Q}{d\omega} \right|_{\substack{\text{Born}\\\text{non-id.}}}
		&=
		\frac{8}{15} \frac{\alpha^3 Z_1^2 Z_2^2 A_Q^2}{\mu ^2} \bigg[10 \sqrt{1-x} 
		 \nonumber\\
		&+ 3 (2-x)\ln \left(\frac{1+ \sqrt{1-x}}{1- \sqrt{1-x}}\right) \bigg] .
	\end{align}
For identical spin-1/2 particles, such as in $e^-e^-$ scattering, with mass $m_{1,2}=m$ and charge $Z_{1,2}=Z$ one finds in agreement with literature~\citep{Fedyushin:1952hg},
      {
\medmuskip=0mu
	\begin{align} \label{eqn:vector_ferm_pp_cs_id}
		\omega \left.\frac{d\sigma_Q}{d\omega} \right|_{\substack{\text{Born}\\ \text{ident.}}  }
		& =
		\frac{4}{15} \frac{\alpha^{3} Z^6}{m^2} \sqrt{1-x} \bigg[17 - \frac{3 x^2}{(2-x)^2} \nonumber \\ 
	& \!\!\!\!\!\!\!	+\frac{12 (2 - x)^4 - 7 (2 - x)^2 x^2 - 3 x^4}{2\left(2-x\right)^3 \sqrt{1-x}} \ln \left(\frac{1+ \sqrt{1-x}}{1- \sqrt{1-x}}\right)\bigg] .
	\end{align}
}%
For identical spin-0 particles, such as in $\alpha$-particle scattering, we obtain in agreement with~\cite{1981PhRvA..23.2851G},
{
\medmuskip=0mu
	\begin{align} 
	\label{eqn:vector_scal_pp_cs_id}
		\omega \left.\frac{d\sigma_Q}{d\omega} \right|_{\substack{\text{Born}\\ \text{ident.}}  }
		& =
		\frac{4}{15} \frac{\alpha^{3} Z^6}{m^2} \sqrt{1-x} \bigg[26 + \frac{6 x^2}{(2-x)^2} \nonumber \\ 
	& \!\!\!\!\!\!\!	+\frac{6 (2 - x)^4 + 7 (2 - x)^2 x^2 + 3 x^4}{\left(2-x\right)^3 \sqrt{1-x}} \ln \left(\frac{1+ \sqrt{1-x}}{1- \sqrt{1-x}}\right)\bigg] .
	\end{align}
}%
We have verified that we obtain the cross sections above also by integrating the squared matrix elements presented in App.~\ref{sec:mat_elements}, which were obtained from a tree-level quantum field theory calculation for $|Z_1|/m_1 = |Z_2|/m_2$. 
The Born cross sections \eqref{eqn:cs_vector_pa}--\eqref{eqn:vector_ferm_pp_cs_id} are shown as dash double-dotted gray lines in Figs.~\ref{fig:sigma_diff} and~\ref{fig:sigma_diff_repulsive}. 

In the Elwert prescription~\citep{Elwert:1939km} mentioned in the introduction, the Born cross sections are multiplied by a ratio of the Sommerfeld factors $S_i$ and $S_f$, defined in Eq.~\eqref{eqn:sommerfeld_enhancement_factor} and which also multiply the exact expressions \eqref{eqn:full_quantum_cs}, \eqref{eqn:dist_cross_section} and \eqref{eq:quad_cs_full_id}.%
\footnote{An extensive physical reasoning why the correction factor involves the {\it ratio} of wave-functions at the origin, $|\chi^{(-)}_{\vec{p}_f}(0)|^2/|\chi^{(+)}_{\vec{p}_i}(0)|^2$,
can be found in~\cite{gould2006electromagnetic}.}
The Elwert cross section is then given by,
\begin{align}
\label{eq:Elwert}
    \omega \left. \frac{d\sigma_I}{d\omega} \right|_{\text{Elwert}} &=
    e^{-2\pi \nu_i}\frac{S_f}{S_i} \omega \left.\frac{d\sigma_I}{d\omega} \right|_{\text{Born}} .
\end{align}
Note that the prefactor  $e^{-2\pi \nu_i}S_f/S_i$ is always greater than unity for attractive and less than unity for repulsive interactions.
Results based on the prescription (\ref{eq:Elwert}) are shown by the dashed red lines in all figures. 
In Figs.~\ref{fig:sigma_diff} and \ref{fig:sigma_diff_repulsive}, one can see that the Elwert cross section deviates from the differential Born cross section towards the kinematic endpoint, \textit{i.e.}~for $x\to 1$. 
In the soft photon region, both expressions give the same result.

\subsection{Quasi-classical Limit} \label{sec:classical_limit}

Classical results from quantum mechanics are obtained by taking the limit $\hbar \to 0$. Let us hence, momentarily, reinstate the factors of $\hbar$ in the Sommerfeld parameter and in the fraction of CM energy carried away by the photon, 
\begin{align}
 \nu_{i,f} = \frac{Z_1 Z_2 \alpha}{\hbar v_{i,f}}, \quad x = \frac{\hbar 2 \omega}{\mu v_i^2} .
\end{align}
For $|\nu_i|\gg 1$ it is then said that the particles move on classical trajectories and both, $|\nu_i|\to \infty$ and $|\nu_f| = |\nu_i|/\sqrt{1-x} \to \infty$ when $\hbar \to 0$. Taking the classical limit therefore requires  the limiting behaviour of the
hypergeometric functions $ F(z) = {}_2F_1(i\nu_f,i\nu_i;1;z) $ and its derivative $F'(z)$ for two large imaginary parameters $\pm i|\nu_{i,f}|$; the positive sign is for attractive and the negative sign for repulsive interactions.

In turn, the limit $x\to 0$ when $\hbar \to 0$ is the long-wavelength (soft photon) limit, implying vanishing recoil of the emitting particle pair. We point out that also $|z|$, the argument of $F,\, F'$, becomes large and $z\to -\infty$ for $x\to 0$. The asymptotic expansion of the hypergeometric functions hence depends on  which of its arguments $|\nu_{i,f}|$ or $|z|$ grows quicker in magnitude. To this end, note that the product of $x$ and $\nu_i$ is independent of $\hbar$, \begin{align}
\label{eq:kappa}
 \kappa \equiv \frac{|\nu_i| x}{2} = \frac{|Z_1 Z_2| \alpha \omega}{\mu v_i^3},
\end{align}
so that the value of $\kappa$ becomes the parameter delineating the possible asymptotic limits~\citep{berestetskii1982quantum}. In the following, we will treat the case $\kappa \gg 1$ observing that it remains compatible with the quasi-classicality condition $x\ll 1$. For the attractive case, such limit in the hypergeometric functions can be calculated with the saddle-point method. A discussion of it can be found in the book by~\cite{Sommerfeld_Spektrallinien2}, which we also follow.  A discussion of the case $\kappa \ll 1$ will be relegated to Sec.~\ref{sec:softphoton}.

We start with attractive interactions, $Z_1Z_2<0$,  and note that for taking the limit $\nu_{i,f}\gg 1$, the contour integral representation of the hypergeometric
functions can be used to rewrite $F(z)$ and $F'(z)$ in the form 
\begin{subequations}
\begin{align} \label{eqn:contour_int}
	F^{(\prime)}(z)&=\frac{e^{-\pi \nu_i}}{2 \pi i} \oint_{0^-}^{1^+} du \, \overset{\scriptscriptstyle (\sim)}{g}(u)
 \, e^{\nu_f f(u)}\,, \\
	f(u)&=
	i\ln\left[\frac{u^y}{(1-u)^y(1-uz)}\right] \,,
\end{align}
\end{subequations}
with $g(u)=1/u$ for $F$ and $\tilde g(u)=i\nu_f/(1-uz)$ for $F'$. The contour in \eqref{eqn:contour_int} circles the singular points $u=0$ and $u=1$ in positive direction. The saddle points are situated at
\begin{align}
	u_0 = \frac{1-y}{2} \left(1+i \cot \frac{\theta}{2}\right)\, .
\end{align}
We can then expand around $u_0$, taking into account the second and third order in the expansion of $f(u)$ yielding for $\nu_i\gg 1$,
\begin{subequations}\label{eqn:FFP_large_nu}
{
\medmuskip=0mu
\begin{align}\label{eqn:F_large_nu_third}
	 F(z) &\approx
	-\frac{e^{-\pi \nu_i+ \nu_f f(u_0)}}{2\pi i u_0 \nu_f^{1/3}}
	\left[
	\mathcal I_1
	-
	\frac{\mathcal I_2}{u_0 \nu_f^{1/3}}
	\right] \,, \\
	\label{eqn:Fp_large_nu_third}
	 F'(z) &\approx
	-\frac{\nu_f e^{-\pi \nu_i+ \nu_f f(u_0)}}{2\pi(1- z u_0 ) \nu_f^{1/3}}
	\left[
	\mathcal I_1
	+
	\frac{z \, \mathcal I_2}{(1- z u_0 ) \nu_f^{1/3}}	\right] \,,
\end{align}
}%
\end{subequations}
where the first integral is the leading order expansion in $\overset{\scriptscriptstyle (\sim)}{g}(u)$  around the saddle point and the second integral is the next to leading order expansion.
The integrals can be identified as the integral representations of the modified Bessel functions
{\medmuskip=0mu
\thinmuskip=0mu
\thickmuskip=0mu
\nulldelimiterspace=1pt
\scriptspace=0pt
\begin{align}
\begin{split} \label{eqn:I1}
	\mathcal I_1 &=
	\int_{-\infty}^{\infty} dx \; e^{-a x^2 + i b x^3} 
	= 
	\frac{2}{3 \sqrt{3} } \frac{a }{b}
	e^{\frac{2}{27}\frac{a^3}{b^2}}
	K_{1/3}\left(\frac{2}{27}\frac{a^3}{b^2}\right)
	\,,
\end{split}	\\
\begin{split} \label{eqn:I2}
	\mathcal I_2 &=
	\int_{-\infty}^{\infty} dx \;x\; e^{-a x^2 + i b x^3} \\
	&=
	\frac{2 i}{9 \sqrt{3} } \frac{a^2 }{b^2}
	e^{\frac{2}{27}\frac{a^3}{b^2}}
	\left[
	K_{1/3}\left(\frac{2}{27}\frac{a^3}{b^2}\right)-
	K_{2/3}\left(\frac{2}{27}\frac{a^3}{b^2}\right)
	\right]
	\,,
\end{split}
\end{align}}%
with $a=-\nu_f^{1/3} f''(u_0)/2$, $b=- i \, f'''(u_0)/6$. Plugging \eqref{eqn:FFP_large_nu} into \eqref{eqn:full_quantum_cs} implies evaluating $\mathcal I_1$ and $\mathcal I_2$ at $\theta=\pi$, which yields in agreement with~\cite{berestetskii1982quantum},
\begin{equation} \label{sigma_dipole_large_nu_att}
	\left.	\omega \frac{d\sigma_{D}}{d\omega} \right|_{\text{classical}} =
		\frac{16 \pi }{3 \sqrt{3}} \frac{\alpha^3 Z_1^2 Z_2^2 A_D^2}{\mu^2 v_i^2} \quad (Z_1Z_2 <0, \, \kappa\gg 1).
\end{equation}
This expression is generally used in the definition of a Gaunt factor for dipole radiation, see Sec.~\ref{sec:gaunt}, and it is related to the original result by~\cite{Kramers:1923} on the inverse process, namely, the classical rate for photon absorption.  
For repulsive interactions, $Z_1Z_2>0$, Eq.~(\ref{sigma_dipole_large_nu_att}) gets multiplied with an overall factor $\exp(-2\pi \kappa)$.

For quadrupole bremsstrahlung, we only possess closed solutions for the double differential cross section in $\omega$ and $\cos\theta$, \textit{i.e.}~\eqref{eqn:dist_cross_section} and \eqref{eq:quad_cs_full_id}.
Therefore, we have to keep the $\theta$-dependence in $\mathcal I_1$ and $\mathcal I_2$ and integrate to get the differential cross section in $\omega$. For the solution in an attractive potential, it can be shown that for $\nu_i \to \infty$, the double differential cross section goes to zero for all values of $\theta$ except for $\theta=\pi$. A tedious calculation, which involves expanding $\mathcal I_1$ and $\mathcal I_2$ around $\theta=\pi$ up to fourth order and integrating the result over $\theta$, yields for quadrupole radiation,
\begin{align}  \label{sigma_quad_large_nu_att}
	\left.	\omega \frac{d\sigma_{Q}}{d\omega}\right|_{\text{classical}}  = 
		\frac{32 \pi^{3/2}}{3^{7/6} \Gamma(1/6)} &\frac{\alpha^3 Z_1^2 Z_2^2 A_Q^2}{\mu^2} \left(\frac{\alpha |Z_1 Z_2| \omega}{\mu v_i^3}\right)^{2/3} \nonumber \\
		&\,\,\,\, (Z_1Z_2 <0, \, \kappa\gg 1).
\end{align}
Again, the repulsive case, $Z_1Z_2 >0$, is obtained by multiplying the right hand side by $\exp(-2\pi \kappa)$.
It is worth noting at this point, that the cross sections for scattering of identical and non-identical particles yield the same classical limit \eqref{sigma_quad_large_nu_att} as they should.%
\footnote{$\Gamma(1/6)\approx 5.5663$ can be related to a complete elliptic integral of the first kind, but no simple relation to resolve $\Gamma(1/6)$ exists.}

Appendix~\ref{sec:classical_cs} presents the classical calculations for energy loss. For dipole radiation, the result for $d\sigma/d\omega$ is integrable to Hankel functions for arbitrary $\kappa$ . When an expansion is made for $\kappa\gg 1$ and the classical velocity is identified with $v_i$, Eq.~(\ref{sigma_dipole_large_nu_att}) follows.%
\footnote{The classical calculation does not resolve the difference between $v_i$ and $v_f$ so that the agreement is in that sense accidental~\cite{berestetskii1982quantum}.}
For the classical treatise of quadrupole radiation, we are able to cast the result on $d\sigma/d\omega$ as an integral over Airy functions, as shown in App.~\ref{sec:classical_Airy}, which can be solved analytically and yields Eq.~\eqref{sigma_quad_large_nu_att} to leading order for $\kappa \gg 1$ after the classical velocity is again identified with $v_i$. 
The asymptotic forms can be seen in the right panels of  Figs.~\ref{fig:sigma_diff} and \ref{fig:sigma_diff_repulsive} where the thin dashed blue lines show the ``hard photon'' ($\kappa\gg1$) expressions \eqref{sigma_dipole_large_nu_att} and~\eqref{sigma_quad_large_nu_att}. In addition, the blue dash-dotted lines are the full classical results for general $\kappa$ derived in App.~\ref{sec:classical_cs}.

\section{Soft-photon limit}
\label{sec:softphoton}

In this section, we comment on the case of soft photon emission, \textit{i.e.},~the limit $x\ll 1$, focusing on spin-1/2 particles as the case of primary interest.
It is long known~\citep{Low:1958sn} that in the soft-photon limit the double-differential emission cross section can be written as the product of elastic scattering cross section $\left.d \sigma/d\cos\theta\right|_{\vec v_i \to \vec v_f}$ and an overall factor $\mathcal{A}_I $ describing the emission (see \textit{e.g.}~\cite{berestetskii1982quantum}),
\begin{align} \label{eqn:soft_cs}
    \left.\omega \frac{d \sigma_I}{d\omega d\cos\theta}\right|_\text{soft}= \mathcal A_I \times  \left.\frac{d \sigma}{d\cos\theta}\right|_{\vec v_i \to \vec v_f} 
     .
\end{align}
$\mathcal{A}_I $ is assembled from the soft emission factors $\propto e_n (p_n \cdot \epsilon^*)/(p_n\cdot q)$ that multiply the individual amplitudes where a photon with four-momentum $q=(\omega, \vec q)$ and polarization vector $\epsilon^*$ is attached to an external leg with four-momentum $p_n$ and charge $e_n$. For the polarization-summed squared matrix element this implies, 
\begin{align} \label{eqn:emission_piece}
\mathcal A_I = -\frac{\omega^2}{(2\pi)^2}\sum_{m,n} \eta_m \eta_n \, e_m e_n \frac{p_m \cdot p_n}{(p_m \cdot q)(p_n \cdot q)}.
\end{align}
Here, the sum runs over all initial (final) states with $\eta = \pm 1$. The analytic structure of~\eqref{eqn:emission_piece} is independent of the spin of the emitting particles~\citep{Weinberg:1965rz} and underlies the theorem implying the cancellation of infrared divergences in quantum electrodynamics, and along analogous lines, with a modified version of~\eqref{eqn:emission_piece}, the cancellation of infrared divergences in the emission of soft gravitons~\citep{Weinberg:1965nx}.

In the dipole case, it is then straight-forward to show that in the non-relativistic expansion, $\mathcal{A}_D$ becomes the squared difference of initial and final state velocity,
\begin{align} \label{eqn:emission_dipole}
\mathcal A_D(\omega, \cos\theta)
&=\frac{2 \alpha A_D^2}{3 \pi} |\vec v_i-\vec v_f|^2 ,
\end{align}
and which is to be multiplied by the Rutherford cross section
\begin{align}\label{eqn:elastic_eI}
  \left.\frac{d\sigma}{d\cos\theta}\right|_{\vec v_i \to \vec v_f}&=
  \frac{8 \pi \alpha^2 Z_1^2 Z_2^2}{\mu^2}\frac{1}{|\vec v_i-\vec v_f|^4} .
\end{align}
It is important to note that~\eqref{eqn:elastic_eI} is {\em exact} in the Coulomb interaction of the scattering particles and also coincides with the Born result.

Integrating the product \eqref{eqn:emission_dipole} with \eqref{eqn:elastic_eI} over the scattering angle yields to leading order in $x$ the Born cross section~\eqref{eqn:cs_vector_pa} after the latter is expanded in $x$ as well%
\footnote{In fact, the full Born cross section could be obtained when we use $v_f = v_i\sqrt{1-x}$ and multiply the factorized differential cross section with $v_f/v_i$ before integrating over the scattering angle; this factor usually appears as a phase space factor in the definition of the cross section but is unity for elastic scattering in the CM frame. 
The agreement is somewhat accidental and beyond the accuracy of the factorization. It has to do with the fact that for  non-relativistic bremsstrahlung the emitted momentum $|\vec q| \leq \mu v_i^2/2$ is negligible with respect to the typical exchanged momentum $|\vec k|\sim \mu v_i$.},
\begin{align}
     \label{eqn:cs_vector_pa_soft}
    \omega \left. \frac{d\sigma_D}{d\omega} \right|_{\text{Born}}^{\text{soft}}
    &= \frac{16}{3} \frac{\alpha^{3} Z_1^2 Z_2^2 A_D^2}{\mu^2 v_i^2}
     \ln \left(\frac{2\mu v_i^2}{\omega}\right) .
\end{align}
Since the result is a direct consequence of his soft-photon theorem~\citep{Weinberg:1965nx} and {\em per se} an exact result in the limit $x\to 0$, \cite{Weinberg:2019mai} recently investigated the validity of~\eqref{eqn:cs_vector_pa_soft} away from $|\nu_i|\ll 1$ and observed that when the scattering approaches the forward direction in~\eqref{eqn:elastic_eI}, the requirement on the smallness of $x$ becomes increasingly stringent for~\eqref{eqn:cs_vector_pa_soft} to hold. A formula was therefore suggested that splits the angular integration into two regimes, where in the forward direction, for $\theta < \theta_c$, a correction factor $\pi^2 \nu_i^2/\sinh^2{\pi \nu_i}$ is applied to the product of \eqref{eqn:emission_dipole} and \eqref{eqn:elastic_eI}. The factor was obtained from an asymptotic analysis, and although not explicitly stated in~\cite{Weinberg:2019mai}, it is just the product of Sommerfeld factors $S_i S_f$  as they appear in the exact formula~\eqref{eqn:full_quantum_cs} in the limit $x\to 0$. In this treatment, $\theta_c$ is a critical angle that can be determined from matching onto the classical limit ($|\nu_i|\gg 1$) for soft photon emission ($\kappa\ll 1$)~\citep{landau1975classical},%
\footnote{The $\theta_c$ dependence drops out in the Born limit, and using the asymptotic classical result for the matching such that $\theta_c$ becomes independent of $\nu_i$ makes the relation to the exact result approximate; 
in actual physical applications Debye screening of the Coulomb collision may need to be taken into account. In Sec.~\ref{sec:thermal_effects} we show that this is only the case for when $\omega\lesssim \omega_p$ where $\omega_p$ is the plasma frequency.}
\begin{equation} \label{sigma_dipole_large_nu_att_soft}
	\left.	\omega \frac{d\sigma_{D}}{d\omega} \right|_{\text{classical}}^{\text{soft}} =
		\frac{16}{3} \frac{\alpha^3 Z_1^2 Z_2^2 A_D^2}{\mu^2 v_i^2}
		\ln\left(\frac{2\mu v_i^3 \, e^{-\gamma}}{\alpha |Z_1 Z_2| \omega}\right),
\end{equation}
where $\gamma\approx 0.5772$ is the Euler-Mascheroni constant.
Such formulation has the benefit of providing an approximate formula for {\em soft} dipole bremsstrahlung for arbitrary~$\nu_i$.
Note that \eqref{sigma_dipole_large_nu_att_soft} is also  obtained by expanding the classical expression \eqref{eqn:classical_cs_full} for small~$\kappa$.

If we now turn to the case of quadrupole emission, the elastic Coulomb scattering cross section for identical spin-1/2 particles of charge $Z$ and mass $m$ reads~\citep{berestetskii1982quantum}
\begin{align}\label{eqn:elastic_ee}
  \left.\frac{d\sigma}{d\cos\theta}\right|_{\vec v_i \to \vec v_f}^\text{ident.}&=
  \frac{32 \pi \alpha^2 Z^4}{m^2}\left[\frac{1}{|\vec v_i-\vec v_f|^4}+\frac{1}{|\vec v_i+\vec v_f|^4}\right.
  \nonumber \\
  & \!\!\! \!\!\! \!\!\! \!\!\!\!\!\!  \left.-\frac{1}{|\vec v_i+\vec v_f|^2|\vec v_i-\vec v_f|^2} \cos\left(\nu_i \ln\frac{|\vec v_i-\vec v_f|^2}{|\vec v_i+\vec v_f|^2}\right) \right].
\end{align}
Since the formula is exact in the Coulomb interaction of the scattering  particles, it is natural to attempt to establish an analytical formula for soft photon emission that is valid for arbitrary $\nu_i$ by following Weinberg's treatment for dipole radiation~\citep{Weinberg:2019mai}. For this, however, there are two obstacles to overcome: one is related to the complicated analytical structure of the cosine-term in~\eqref{eqn:elastic_ee} and another is related to the fact, that a naive application of the factorization~\eqref{eqn:soft_cs} yields a prohibitive requirement on the smallness of~$x$. The rest of this section will be concerned with showing how we can overcome these two obstacles, and we start with the latter. The final formula is presented in~\eqref{eqn:quad_emission_weinberg}.

\begin{figure}[t]
	\includegraphics[width=\columnwidth]{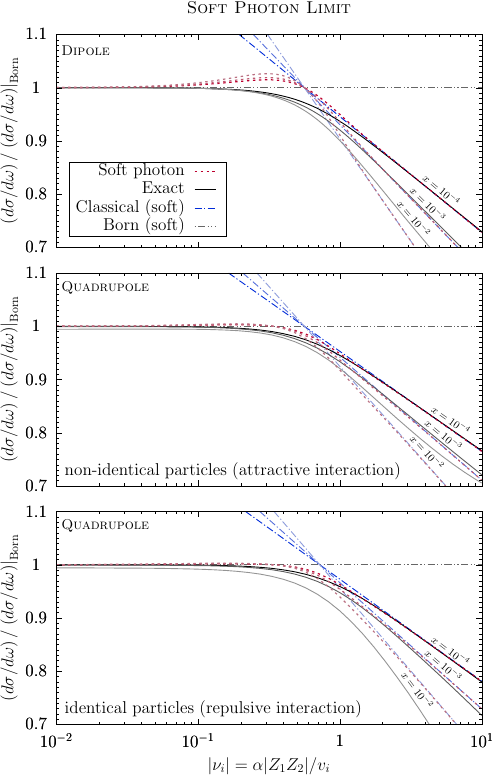}%
	\caption{Differential cross section $\omega\, d\sigma/d\omega$ evaluated at $x=10^{-4},10^{-3},10^{-2}$ (darker to lighter lines) as a function of $\nu_i$ for dipole emission (top) and quadrupole emission for attractive interactions (center) and for identical particles (bottom).
	The general soft photon cross sections, \textit{i.e.},~(37) in \cite{Weinberg:2019mai} and~\eqref{eqn:quad_emission_weinberg} of this work, with $\theta_c$ matched to the soft classical limit, are shown by the dashed red lines. They are compared to
	the classical limits~\eqref{sigma_dipole_large_nu_att_soft} and~\eqref{sigma_quad_large_nu_att_soft} for $\kappa\ll1$ in dash-dotted blue, the Born limits~\eqref{eqn:cs_vector_pa_soft},~\eqref{eqn:vector_ferm_pp_cs_soft} and~\eqref{eqn:vector_ferm_pp_cs_id_soft} in dash double-dotted gray and the exact results~\eqref{eqn:full_quantum_cs} and angle-integrated~\eqref{eqn:dist_cross_section} and~\eqref{eq:quad_cs_full_id} in black.
	\label{fig:soft_limit}}
\end{figure}

For quadrupole radiation, it turns out that although using~\eqref{eqn:emission_piece} for the multiplying factor $\mathcal{A}_Q$ yields the correct asymptotic behavior for $x\to 0$, its range of applicability will be exceedingly small. A better result can be obtained by correcting for the fact that different momenta (by an amount $\pm q$)  are exchanged in the individual diagrams that mediate the Coulomb interaction; in the Born limit, this is best seen from the diagrams of Fig.~\ref{fig:feyman_diag}. Correcting each term in~\eqref{eqn:emission_piece} for the actual exchanged momentum $k_m$ in the emission from an external leg of momentum $p_m$ by multiplying it with the factor
${k^4_\text{elastic}}/(k^2_m k^2_n)$, where $k_\text{elastic}$ denotes the exchanged momentum for $q\to 0$, yields
\begin{align} \label{eqn:emission_piece_quad}
\mathcal A_Q^\text{corr} 
 &= \frac{2\alpha A_Q^2}{15\pi}\left[4v_i^4+4v_f^4-v_i^2v_f^2 \left(5+3 \cos^2\theta\right)\right], \nonumber \\
&=\frac{2\alpha v_i^4 A_Q^2}{15\pi}\left[3(1-\cos^2\theta) -3x(1-\cos^2\theta)+4x^2\right].
\end{align}
The difference with respect to using~\eqref{eqn:emission_piece} is in the quadratic term $4x^2$ which would otherwise have read~$x^2$.

The importance of using~\eqref{eqn:emission_piece_quad} over an emission factor  obtained from~\eqref{eqn:emission_piece} is exemplified when
  considering the Born regime $|\nu_i|\ll 1$ for which one may take the cosine in the last term of~\eqref{eqn:elastic_ee} to unity.  Integrating the product of~\eqref{eqn:elastic_ee} with~\eqref{eqn:emission_piece_quad} over the scattering angle one then obtains to leading order in $x$,
\begin{align} 
 \label{eqn:vector_ferm_pp_cs_id_soft}
		\omega \left.\frac{d\sigma_Q}{d\omega} \right|_{\substack{\text{Born}\\\text{ident.}}  }^{\text{soft}}
		\!\!\!
		& =
		\frac{4}{15} \frac{\alpha^{3} Z^6}{m^2} \bigg[17 +12 \ln \left(\frac{2\mu v_i^2}{\omega}\right)\bigg] .
	\end{align}
This expression now not only agrees in the leading logarithmic term with the result that is obtained from the Born cross section~\eqref{eqn:vector_ferm_pp_cs_id} after an expansion in $x$ has been performed, but also in the constant coefficient~17.
For example, while for dipole radiation the Born cross sections~\eqref{eqn:cs_vector_pa} and~\eqref{eqn:cs_vector_pa_soft} agree to better than one part in $10^{3}$ for $x=10^{-2}$, the cross section obtained from \eqref{eqn:emission_piece} for quadrupole radiation would deviate by 25\%  from~\eqref{eqn:vector_ferm_pp_cs_id} and the error would drop to $10^{-3}$ only at an unreasonable  exponentially smaller value~$x=10^{-1000}$.
Using instead \eqref{eqn:emission_piece_quad} the agreement between~\eqref{eqn:vector_ferm_pp_cs_id_soft} and~\eqref{eqn:vector_ferm_pp_cs_id} is as good as for dipole radiation. 
Similar conclusions hold for non-idential particles, and for completeness we record the soft-photon limit of the quadrupole emission cross section for non-identical particles which is obtained by integrating the Rutherford cross section~\eqref{eqn:elastic_eI} with~\eqref{eqn:emission_piece_quad} and which in leading order in~$x$ reads,
\begin{align}
\label{eqn:vector_ferm_pp_cs_soft}
		\omega \left.\frac{d\sigma_Q}{d\omega} \right|_{\substack{\text{Born}\\\text{non-id.}}}^{\text{soft}}
		\!\!\!
		&=
		\frac{16}{15} \frac{\alpha^3 Z_1^2 Z_2^2 A_Q^2}{\mu ^2} \bigg[5
	+ 3 \ln \left(\frac{2\mu v_i^2}{\omega}\right) \bigg] .
	\end{align}

Equipped with the appropriate factors~\eqref{eqn:elastic_ee} and~\eqref{eqn:emission_piece_quad}, an analysis like the one performed for dipole radiation by~\cite{Weinberg:2019mai} and as mentioned above may now be performed for the quadrupole case.  The remaining difficulty is that the cosine-term in \eqref{eqn:elastic_ee} cannot be integrated analytically. In the limit $|\nu_i|\ll 1$ this term goes to unity while in the limit $|\nu_i|\gg 1$ it oscillates quickly  so that the interference term vanishes in the integral over $\theta$, and as one expects from the classical limit where the concept of identical particles is not present. 
We overcome the problem caused by the cosine-term  by making the replacement $\cos(\dots)\to \xi = 1(0)$ in the interference term of~\eqref{eqn:elastic_ee} for identical (non-identical) particles. When we then split the $\theta$-integral in the manner of~\cite{Weinberg:2019mai}, it turns out that the interference term for identical particles will still vanish in the classical limit as it should due to the introduced factor $S_iS_f|_{x\to 0}\to 0$ for $|\nu_i|\to \infty$; for non-identical particles it vanishes trivially because $\xi =0$. Conversely, this prescription also retains the correct form in the Born limit, and hence yields an analytical formula for the soft cross section that has the correct asymptotic forms for $|\nu_i|\ll 1$ and $|\nu_i|\gg 1$ and becomes a reasonable approximation for intermediate values of $\nu_i$, 
\begin{widetext}
\begin{align} \label{eqn:quad_emission_weinberg}
    \left.\omega \frac{d \sigma_{Q}}{d\omega}\right|_{\text{soft}} &=
    \frac{8}{15} \frac{\alpha^{3} Z_1^2 Z_2^2 A_Q^2}{\mu^2}
    \Bigg\{
    13-\frac{6+3\xi}{\zeta^2+1}
    +6\ln \left(\frac{2\mu v_i^2}{\omega \zeta}\right)
     + \left. S_i S_f\right|_{x=0}\left[\frac{6+3\xi}{\zeta^2+1}-\frac{6+3\xi}{2}+ 6\ln \zeta
    \right]
    \Bigg\},
\end{align}
\end{widetext}
with $\zeta\equiv \sqrt{(1+\cos\theta_c)/(1-\cos\theta_c)}$.
In the Born limit, $|\nu_i|\ll1$, we have $\left. S_i S_f\right|_{x=0}\to 1$, the $\zeta$ dependence in \eqref{eqn:quad_emission_weinberg} drops out, and we retrieve \eqref{eqn:vector_ferm_pp_cs_id_soft} or \eqref{eqn:vector_ferm_pp_cs_soft}, respectively, depending on the value of $\xi$. In the classical limit, $|\nu_i|\gg 1$, we get $\left. S_i S_f\right|_{x=0}\to 0$ and we can identify $\zeta=|\nu_i| e^{\gamma+1/2}\gg 1$ by matching onto the classical result for soft photon emission ($\kappa\ll 1$),
\begin{equation} \label{sigma_quad_large_nu_att_soft}
	\left.	\omega \frac{d\sigma_{Q}}{d\omega} \right|_{\text{classical}}^{\text{soft}}
	\!\!\!=
		\frac{8}{15} \frac{\alpha^3 Z_1^2 Z_2^2 A_Q^2}{\mu^2}
	\left[10+6
		\ln\left(\frac{2\mu v_i^3 \, e^{-\gamma}}{\alpha |Z_1 Z_2| \omega}\right)
	\right],
\end{equation}
which we obtain by expanding the classical expression \eqref{eqn:eqn:classical_quad_cs} for small~$\kappa$. Note that in the classical limit the $\xi$ dependence in \eqref{eqn:quad_emission_weinberg}  becomes negligible as it should. 
The above equations are valid for attractive and repulsive interactions since the factor $\exp(-2\pi \kappa)$ that gets usually multiplied to the classical cross sections going from $Z_1Z_2 >0$ to $Z_1Z_2 <0$ goes to unity in the soft photon limit $\kappa\ll 1$. Finally we note that one can in fact improve on~\eqref{eqn:quad_emission_weinberg} to extend the validity to larger values of~$x$. A formula for this is presented in App.~\ref{app:approximate_formula}.

The behavior of \eqref{eqn:quad_emission_weinberg} between the Born and classical limits---with $\zeta$ fixed as outlined above---is shown in the middle and bottom panel of Fig.~\ref{fig:soft_limit} for three values of $x$. As can be seen, the soft cross section \eqref{eqn:quad_emission_weinberg} is a good approximation up to values of $x=10^{-2}$ in the Born limit, while in the classical limit there are clearly visible deviations from the exact result due to the stricter condition $\kappa\ll 1$ for the classical result that is used in the matching.
For comparison, the result of \cite{Weinberg:2019mai} for dipole radiation with fixed $\zeta$ is shown in the top panel of Fig.~\ref{fig:soft_limit}.
The soft limits are also shown by the thin dotted gray (blue) lines in the left (right) panels of Figs.~\ref{fig:sigma_diff} and \ref{fig:sigma_diff_repulsive} for the Born (classical) limit.

\section{Effective energy loss}\label{sec:eff_en_loss}

\begin{figure*}[tb]
	\includegraphics[width=0.488\textwidth]{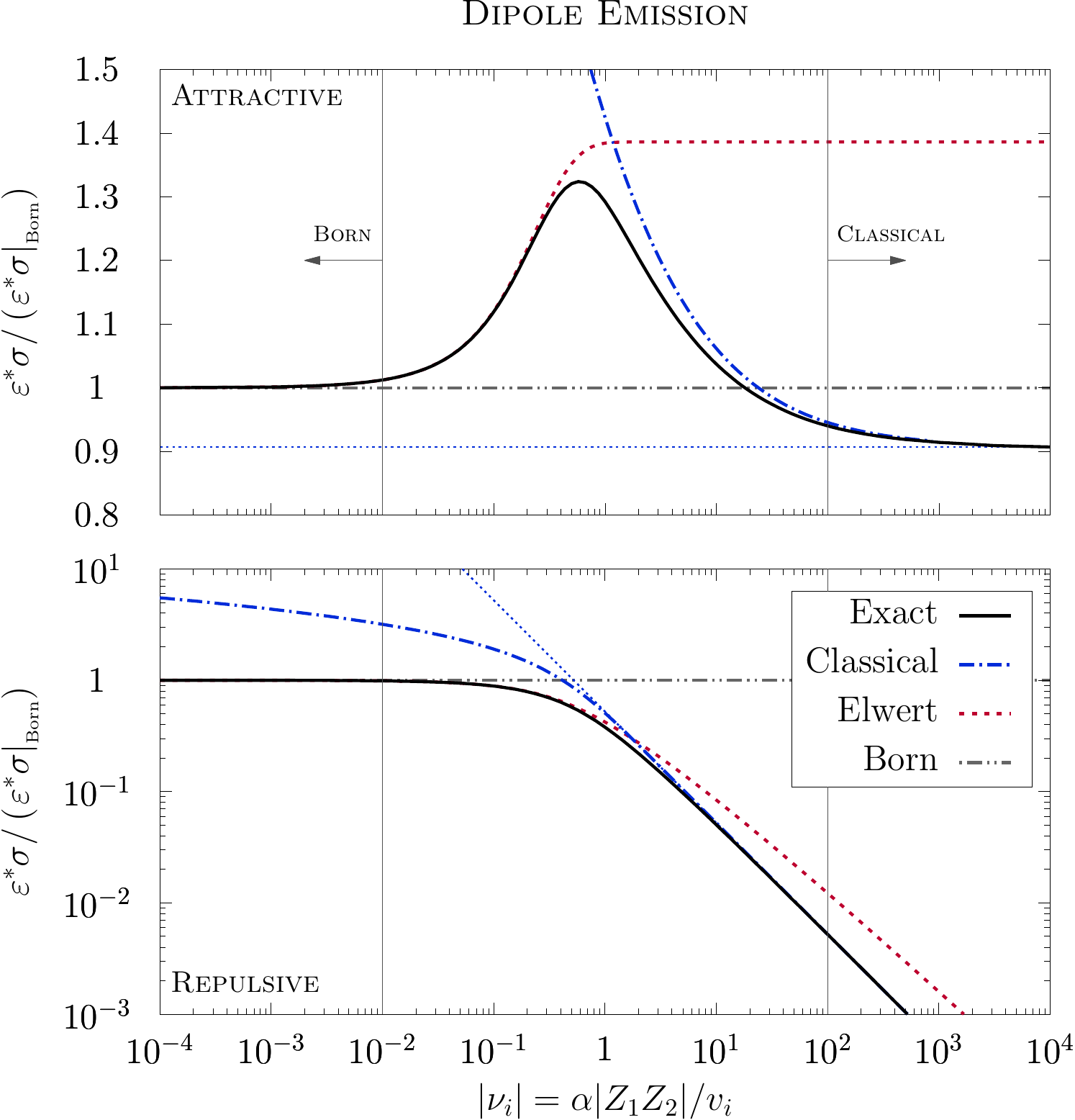}%
	\hfill
	\includegraphics[width=0.48\textwidth]{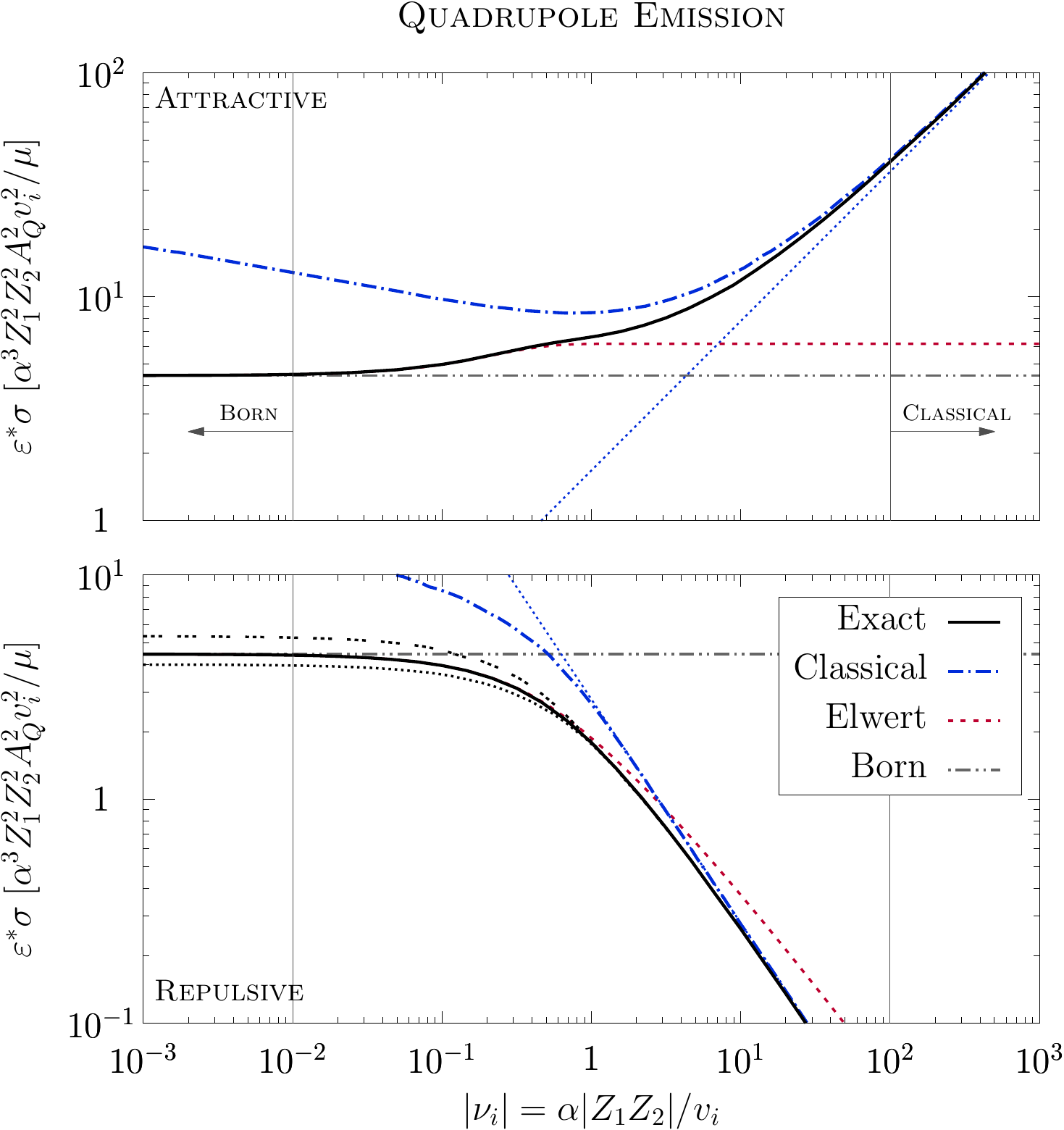}
	\caption{\textit{Left panel:} Effective energy loss from dipole emission, normalized to the Born result [first term of Eq.~\eqref{eqn:Eeff_born} (dash double-dotted gray line)] as a function of $\nu_i$ for an attractive (top) and a repulsive (bottom) interaction. The full non-relativistic quantum mechanical result from Eq.~\eqref{eqn:full_quantum_cs} is shown in solid black. The rigorous classical limit from Eq.~\eqref{eqn:classical_cs_full} containing the Hankel functions is shown in dash-dotted blue, while the asymptotic expressions, \textit{i.e.}~the dipole terms in Eqs.~\eqref{eqn:Eeff_cl_attractive} and \eqref{eqn:Eeff_cl_repulsive}, are shown by the thin dotted blue line. The Elwert approximation is shown in dashed red. \textit{Right panel:} Effective energy loss from quadrupole emission as a function of $\nu_i$ for an attractive (top) and a repulsive (bottom) interaction. The exact result for non-identical particles from Eq.~\eqref{eqn:dist_cross_section} is shown in solid black. The rigorous classical limit from Eq.~\eqref{eqn:eqn:classical_quad_cs} containing the Hankel functions is shown in dash-dotted blue, while the asymptotic expressions, \textit{i.e.}~the quadrupole terms in Eqs.~\eqref{eqn:Eeff_cl_attractive} and \eqref{eqn:Eeff_cl_repulsive}, are shown by the thin dotted blue line. The Elwert approximation is shown in dashed red. For comparison, the effective energy loss for identical particles obtained from Eq.~\eqref{eq:quad_cs_full_id} is shown in dotted black for spin-1/2 and in triple-dotted black for spin-0 particles in the bottom right.
	\label{fig:compare_cross_sections_enhanced}}
\end{figure*}

Since the exact expressions for multipole radiation obtained in Sections~\ref{sec:dipole} and \ref{sec:quadrupole} are rather complicated and sometimes numerically challenging to evaluate, it is of interest to show how the exact solutions behave in between the Born and the classical regimes. We investigate this by comparing the effective energy loss or ``effective retardation'', \textit{i.e.},~the energy-weighted integral over the emission cross section $d\sigma/d\omega$ up to the kinematic endpoint,
\begin{align} \label{eqn:Eeff}
	\EeffQD{I}=\int_{0}^{\frac{\mu v_i^2}{2}} d\omega \; \omega  \frac{d\sigma_I}{d\omega}  \, ,
\end{align}
and which is a quantity of general interest.%
\footnote{
For example, the luminosity in a non-relativistic Max\-wellian gas of such colliding particles with number densities $n_{1,2}$ and common temperature $T$ can be computed from the CM frame expression through~\citep{1984ApJ...280..328D} $ l_I = \frac{n_1 n_2}{(1+\delta_{12})} \sqrt{\frac{2}{\pi}} \left(\frac{\mu}{T}\right)^{3/2}  \int dv_i v_i^3 e^{-\mu v_i^2/2T} \varepsilon^* \sigma_I(v_i) $.
Another example is the energy loss  of electrons in the passage of a gas of heavy ions, for which $-dE/dx \simeq n_I \Eeff$ where $n_I$ is the number density of scattering centers.
}
 
 In the Born limit, the effective energy loss including dipole and quadrupole radiation for non-identical particles is obtained by integrating~\eqref{eqn:cs_vector_pa} and~\eqref{eqn:vector_ferm_pp_cs},
\begin{align} \label{eqn:Eeff_born}
	\left.\Eeff \right|_{\substack{\text{Born}\\\text{non-id.}}}
 &= \frac{\alpha^3 Z_1^2 Z_2^2}{\mu} \left[
	\frac{16}{3}A_{D}^2	 +
	\frac{40}{9} A_{Q}^2	 \, v_i^2
	\right]
	 \,.
\end{align}
The result is valid for both, attractive and repulsive interactions as the Born limit does not distinguish these cases.
For identical particles only quadrupole radiation exists,  $A_D=0$, \lss{$A_Q=Z/2$}. Writing $Z_{1,2}=Z$, $m_{1,2}=m$, and $\mu=m/2$ one obtains from integrating Eqs.~\eqref{eqn:vector_ferm_pp_cs_id} and~\eqref{eqn:vector_scal_pp_cs_id},
\begin{align} \label{eqn:Eeff_born_id}
\left.\EeffQD{Q}\right|_{\substack{\text{Born}\\\text{ident.}}} &=\frac{5}{36}\frac{\alpha^3 Z^6}{m} v_i^2
\times
\begin{cases}
    (44-3\pi^2)  & s=1/2
 \,,\\
    (6\pi^2-40)  & s=0\,.
\end{cases}
\end{align}
As expected, in the Born limit quadrupole emission is suppressed by a factor $v_i^2$ with respect to dipole emission.

Turning now to the classical expressions, from \eqref{sigma_dipole_large_nu_att} and \eqref{sigma_quad_large_nu_att} the effective energy loss for attractive interactions in the classical limit up to quadrupole order reads, 
\begin{align} \label{eqn:Eeff_cl_attractive}
	\left.\Eeff\right|_{\substack{\text{\scriptsize asympt.}\\\text{\scriptsize classical}}} = &\frac{\alpha^3 Z_1^2 Z_2^2}{\mu} \bigg[
	\frac{8\pi}{3\sqrt{3}}A_{D}^2 +
	\frac{8\pi^{3/2} 2^{1/3} }{3^{1/6} 5 \Gamma(1/6)}
	\nonumber \\	 & \times
	|\alpha Z_1 Z_2|^{2/3} A_{Q}^2	 \, v_i^{4/3} 
	\bigg] \quad (Z_1Z_2<0).
	 \,
\end{align}
For repulsive interactions, the energy loss receives equally important contributions from the kinematic regimes $\kappa\gg 1$ and from $\kappa \ll 1$. Hence, using only the $\kappa\gg 1$ limit in Sec.~\ref{sec:classical_limit} would underestimate the result. However, the effective energy loss can be calculated directly from the classical theory using Eq.~\eqref{eqn:classical_EL} and rewriting the time integral to a line integral along the particles' relative distance from infinity to the distance of closest approach and back to infinity (see \textit{e.g.}~\cite{landau1975classical}). One then arrives at
\begin{align} \label{eqn:Eeff_cl_repulsive}
	\left. \Eeff\right|_{\substack{\text{\scriptsize asympt.}\\\text{\scriptsize classical}}} = \frac{8\pi}{9} \frac{\alpha^2 Z_1 Z_2}{\mu}
	& \left[
	 A_{D}^2 v_i	 +
	 A_{Q}^2	 \, v_i^3 
	\right] \nonumber \\ & \quad  \quad\quad \quad(Z_1Z_2>0)
	 \,.
\end{align}
Note that in the attractive case, the dipole term is of the same order in $v_i$ in the Born and in the classical limit while the quadrupole term is only suppressed by a factor $v_i^{4/3}$ w.r.t.~the dipole term in the classical limit as opposed to a factor $v_i^2$ in the Born limit. In the repulsive case,  the quadrupole term is suppressed by a factor $v_i^2$ w.r.t.~the dipole term but the classical effective energy loss shows an overall suppression of $v_i/(\alpha Z_1 Z_2)$ w.r.t.~the Born limit; hence the single powers of $Z_{1,2}$ in \eqref{eqn:Eeff_cl_repulsive}.

We point out again that while the Born approximation distinguishes between identical and non-identical particles, this concept is not present in the classical calculation. On the other hand, the classical approximation distinguishes between attractive and repulsive potentials while---as is well known---the Born approximation does not. The exact calculation presented in Sec.~\ref{sec:quadrupole} captures all of these aspects and interpolates smoothly between the Born and classical limit.

In Fig.~\ref{fig:compare_cross_sections_enhanced} we show the full effective energy loss calculation as obtained from Eqs.~\eqref{eqn:dist_cross_section} and \eqref{eq:quad_cs_full_id} and compare it with the expressions listed above. In addition, we also investigate here the energy loss based on the Elwert prescription~\eqref{eq:Elwert}.
As can be seen, 
$	\left. \Eeff\right|_{\text{\scriptsize Born}} $ is numerically within a few percent of the effective energy loss calculated from the exact cross sections for $|\nu_i| \lesssim 10^{-2}$.
The Elwert prescription extends the range of validity of $	\left. \Eeff\right|_{\text{\scriptsize Born}} $  to $|\nu_i| \lesssim 0.5$. 
The reason is that the ratio of Sommerfeld factors, which get multiplied onto the Born cross section account for the larger (smaller) probability of finding the colliding particles at same position due to the attractive (repulsive) Coulomb interaction and therefore increases (decreases) the cross section accordingly, as detailed in Sec.~\ref{sec:born_limit}. 
For $|\nu_i| \gtrsim 10$ the exact energy loss is well described by the classical approximation derived in App.~\ref{sec:classical_cs} in terms of Hankel functions (shown as blue dash-dotted lines), whose leading order terms in the expansion $|\nu_i| \gg 1$, Eqs.~\eqref{eqn:Eeff_cl_attractive} and \eqref{eqn:Eeff_cl_repulsive}, are shown as blue dotted lines.
For attractive interactions, the classical approximation asymptotes to Eq.~\eqref{eqn:Eeff_cl_attractive} for $|\nu_i| \gtrsim 10^3$ while for repulsive interactions, it already reaches its asymptotic form of  Eq.~\eqref{eqn:Eeff_cl_repulsive} at $|\nu_i| \gtrsim 1$.

While the energy loss in dipole emission for $Z_1Z_2<0$ only deviates from the Born approximation by a maximum $\sim 30 \%$ at a given $\nu_i$, the energy loss in quadrupole emission can deviate from the Born approximation by several orders of magnitude since the classical and Born limit show different scalings with velocity. Note, however, that even though quadrupole radiation in an attractive potential gets enhanced with respect to the Born approximation for large $\nu_i$ while dipole radiation does not, the former can never become larger than the latter since it is additionally suppressed by $v_i^2$ as can be seen in Eq.~\eqref{eqn:Eeff_cl_attractive}. To this end, note that the units on the $y$-axes on the right panel of Fig.~\ref{fig:compare_cross_sections_enhanced} carry an additional $v_i^2$ factor, and the energy loss for attractive interactions in the quadrupole case does not grow without bounds as $\nu_i\to \infty$, but rather vanishes instead.

\section{Electron-Electron Gaunt factor}
\label{sec:gaunt}

The net rate of photon absorption in free-free transitions and its inverse process, bremsstrahlung, is of ample importance in the understanding of opacity and emissivity of astrophysical environments. In the scattering of electrons on ions of charge $Z$ (dipole case) it has become traditional to express the exact rates as a product of the classical electrodynamics result by \cite{Kramers:1923} and a  ``free-free Gaunt factor'' $g_{\rm ff}$. In terms of the bremsstrahlung cross section, with (\ref{sigma_dipole_large_nu_att}) specialized to $Z_1=Z$ $Z_2=-1$, $\mu\simeq m_e$, 
\begin{align}
\label{eq:GauntDipole}
   \omega \frac{d\sigma_D}{d\omega} = 
    \frac{16\pi}{3\sqrt{3}} \frac{\alpha^3 Z^2}{m_e^2 v_i^2}
   \times  g_{\mathrm{ff}}(\omega,v_i) .
\end{align}
Because of the numerical demanding nature of evaluating hypergeometric functions, Gaunt factors are being tabulated.  The classic compilation is by \cite{1961ApJS....6..167K} with literature that continues into the most recent past~\citep{vanHoof:2014bha,Chluba:2019ser}; for soft-photon emission an improved Gaunt factor was recently  proposed by~\cite{Weinberg:2019mai} as discussed in Sec.~\ref{sec:softphoton}.

Having gathered insight into the exact non-relativistic bremsstrahlung cross section in all kinematic regimes, we are now in a position to make an informed proposal for a Gaunt factor for electron-electron bremsstrahlung, $g_{ee}(\omega, v)$. Given the complexity of the result, there are down-sides with any simple definition. For example, staying with historical convention and defining the Gaunt factor through the classical result, one notes that for quadrupole emission in a repulsive potential there are important contributions---as exemplified in the study of effective energy loss in the previous section---from both, ``soft'' and ``hard'' photons, defined through $\kappa\ll 1$ and $\kappa\gg 1$ with $\kappa$ of Eq.~(\ref{eq:kappa}), respectively. One could then define the Gaunt factor as the multiplicative factor that brings the \textit{sum} of classical hard and soft cross sections  for quadrupole radiation (\ref{sigma_quad_large_nu_att}) and (\ref{sigma_quad_large_nu_att_soft}) into agreement with the full result. However, this has the unpalatable property that there is no easy analytical limiting expression for $g_{ee}$.

The kinematic region of largest practical interest is the Born regime. The reason is that for $Z=1$ Coulomb corrections remain comparably smaller than for electron-ion interactions. For example, considering the case of a hot galaxy cluster gas with a ballpark temperature of $T\simeq 10\,{\rm keV}$ and a typical electron velocity of $v \sim \sqrt{2T/m_e} \simeq 0.2$ implies $|\nu_i| \simeq 0.04$, well in the validity regime of the Born cross section. In addition, as we have observed above, the Elwert prescription extends the validity to larger $|\nu_i|$, and we propose a definition of the Gaunt factor that is based on the latter,
\begin{equation}
 \label{eq:GauntQuadrupole}
   \frac{d\sigma_{Q}}{d\omega}  = 
   \frac{\nu_f}{\nu_i}\frac{1-e^{-2\pi \nu_i}}{1-e^{-2\pi \nu_f}}  
  	 \left.\frac{d\sigma_Q}{d\omega} \right|_{\substack{\text{Born}\\ \text{ident.}}  } \times g_{ee}(\omega, v) ,
\end{equation}
where $\left. {d\sigma_Q}/{d\omega} \right|_{\substack{\text{Born}\\ \text{ident.}}  } $ is given through~\eqref{eqn:vector_ferm_pp_cs_id}. The Elwert factor is important to capture the spectrum for hard photons, and for $|\nu_i|\gtrsim 0.1$ it is required to yield the appropriate suppression in the effective energy loss.  In the Born regime, for $|\nu_i|\ll 1$, it then follows that $g_{ee}\to 1$ and $ g_{ee} < 1 $ otherwise; deviations from unity are most important for $|\nu_i| \gtrsim 1$.
Lastly, we note that a tabulation of $g_{ee}$ together with its  thermally averaged version over a large kinematic regime remains a challenging task as it requires a precise evaluation and integration of $F$ and $F'$ for large parameters and argument. This will be presented elsewhere~\citep{inprep2020}.%
\footnote{A previous compilation exists~\citep{2002NCimB.117..359I}, where a quadrupole Gaunt factor was defined as a multiplicative factor of the dipole cross section. In lieu of a full result, the tabulation could only be based on the Elwert prescription and results are hence not accurate for $\nu_i\gg 1$. A thermally averaged version of the  Born cross section~\eqref{eqn:vector_ferm_pp_cs_id} for a Maxwellian gas has been given by~\cite{Maxon1967}.}

\section{In-medium effects}
\label{sec:thermal_effects}

In an actual physical setting, the processes discussed in this paper can be subject to changes by the modified dispersive properties of photons inside media. The principal effects to consider are the screening of the Coulomb interaction in the collision of the two charged particles and the effective final-state photon mass. In this section, we briefly discuss when these effects become important.

As for the emission of a real (transverse) photon, it remains unaltered in a medium for as long as the frequency of the photon exceeds the plasma frequency, $\omega \gg \omega_p$.%
\footnote{For completeness, it should also be mentioned that the in-medium vertex renormalization constant $Z_T$ that multiplies the photon emission amplitude remains close to unity~\citep{Raffelt:1996wa}.}
In an ionized non-relativistic and non-degenerate plasma of temperature $T$---a case of ample interest in the astrophysical context---the plasma frequency to leading order in $T/m_e$ is given by~\citep{Raffelt:1996wa},
\begin{align}
   \omega_p \simeq \sqrt{ \frac{4\pi\alpha n_e}{m_e}} \simeq 10^{-11}\,{\rm eV} \, \left(\frac{n_e}{0.1\, \rm cm^{-3}} \right)^{1/2} . 
\end{align}
In the second relation we have normalized to an electron density as it may, {\it e.g.}, be found in the cores of galaxy clusters~\citep{Sarazin:1986zz}. We  hence conclude that the emission itself remains unaffected in the post-recombination Universe in the observable part of the spectrum for as long we do not enter the vicinity or interior of stellar objects; in the latter case they are, of course, of critical importance.  For example, the  electron density in the solar corona is $n_e\lesssim 10^9\,{\rm cm^{-3}}$, affecting the radio emission at 100~MHz and below~\citep{2018A&A...614A..54V}.

In the non-relativistic collision of two particles, their electrostatic interaction is screened for three-momentum transfers $|\vec k| \lesssim k_D$ where $k_D$ is the Debye screening scale.%
\footnote{In Coulomb gauge the 00-component of the photon propagator is dotted into the velocity unsuppressed temporal parts of the (external) currents. Inside an isotropic medium, going from zero to non-zero temperature amounts to a replacement of $k^{-2}$ by $[k^2 - \Pi_L(k^0, |\vec k|) ]^{-1}$  where $\Pi_L$ is the longitudinal part of the photon self-energy;  the static limit yields the Debye screening scale, $k_D^2 \equiv \Pi_L(0,|\vec k|)$ and in the case of interest, energy-exchange can indeed be neglected, $k^0\ll |\vec k|$.}
In turn, the emission of a photon of energy $\omega$ requires a minimum momentum transfer $k_{\rm min}(\omega) $ from the 3-body phase space, and for $ k_{\rm min}(\omega) \gg  k_D $ screening plays no role. This can be turned into a condition on the photon energy,  
\begin{align}
\label{debye}
  \omega \gg \frac{\mu v_i^2}{2}\left[1- \left(1-\frac{k_D}{\mu v_i}\right)^2  \right] \simeq  v_i k_D ,
\end{align}
where we have expanded in the second relation on the account that $k_D \ll \mu \langle v_i \rangle$; 
for the canonical case, $k_D^2 = 4\pi \alpha T^{-1} \sum_i n_i Z_i^2 $ \citep{Raffelt:1996wa}
where the sum is over all  particles with charge $Z_i$ (electrons and ions). Given that $\omega_p /k_D \sim \sqrt{T/m_e} \ll 1$, as per our premises, we note that Eq.~\eqref{debye} or $\omega \gg \langle v_i \rangle k_D $ yields a similar requirement as~$\omega\gg \omega_p$.
For example, in a hot cluster gas [$n_e \simeq 10^{-3}\,{\rm cm^{-3}}, \, T \simeq 10\,\rm keV$~\citep{1966ApJ...146..955F}] the requirement~\eqref{debye} reads $\omega\gg 10^{-12}\,{\rm eV}$ and inside HII regions [$n_e \simeq 10\,{\rm cm^{-3}}, \, T \simeq 1\,\rm eV$,~\citep{1965ApJ...142..135T}] the above requirement reads $\omega\gg 10^{-10}\,{\rm eV}$. We therefore conclude that screening is of little relevance when considering bremsstrahlung in dilute astrophysical plasmas.

Finally, in passing, we also comment on the possibility where multiple Coulomb-collisions between neighboring particles modify the scattering probability with the general effect of reducing the cross section for bremsstrahlung \citep{Landau:1953um,Migdal:1956tc}. The maximum coherence length that is inherent to the process can be taken as $\lambda_{\rm coh} \sim 1/k_D$, which we may compare to the electrons' mean free path $\lambda_{\rm mfp} \sim 1/(\sigma_t n)$. Here $\sigma_t = 4\pi\alpha^2 Z^2/(m_e^2 v_i^4 ) \ln\Lambda_{\rm Coul}$ is the non-relativistic transport cross section between electrons and---for the sake of the argument---some population of ions with charge $Z$ and number density $n\sim n_e$; $\ln \Lambda_{\rm Coul}$ is the Coulomb-logarithm. Taking the ratio $\lambda_{\rm coh} /\lambda_{\rm mfp} \sim 10^{-17}Z^2\ln\Lambda_{\rm Coul} \, (n_e/0.1\ {\rm cm^{-3}})^{1/2} (10\ {\rm keV}/T)^{3/2}  $ demonstrates that such collective effects can  generally be considered as absent in dilute astrophysical gases. In the cases when it actually becomes relevant, stricter figures of merit than $\lambda_{\rm mfp}$ should  be used~\citep{berestetskii1982quantum}; for a general textbook discussion on the above effects, see {\it e.g.}~\cite{Raffelt:1996wa}.

It goes without saying that the immense diversity of physical conditions that are of astrophysical interest implies that medium effects, when they become of importance,
 require the dedicated study on a case-by-case basis. Here, we have demonstrated that regarding the non-relativistic bremsstrahlung process in the post-recombination dilute interstellar medium these effects can largely be  neglected.

\section{Conclusions}
\label{sec:conclusions}

In this work we present the full quantum mechanical treatment for the quadrupolar single-photon emission which is exact in the low energy scattering of two electrically charged spin-0 or  spin-1/2 particles. The double differential cross sections in the center-of-mass scattering angle $\theta$ and photon energy $\omega$ is presented in Eq.~\eqref{eqn:dist_cross_section} for non-identical and in Eq.~\eqref{eq:quad_cs_full_id} for identical particles. The latter formula applies to the important case of the scattering of a pair of electrons for which quadrupole radiation is the leading energy loss  process.

For dipole radiation, the result can be cast into the rather short form~\eqref{eqn:full_quantum_cs} by making use of the differential equation for the hypergeometric function. This is not possible anymore for the quadrupole case. The analytic structure of the result is considerably more complex. Its building blocks are the elements of a tensor of Coulomb transition matrix elements, Eq.~\eqref{eq:QmnDef}, and we are able to obtain the elements in closed form by making use of an integral formula for confluent hypergeometric functions that was established by Nordsieck, see~\eqref{QinNordsieck} and \eqref{eqn:nordsieck_integral}.

The results apply to all non-relativistic kinematic regimes and yield the correct Born and quasi-classical limits for $|\nu_i| \ll 1 $ and $|\nu_i| \gg 1 $, respectively. We show how these limits are derived from the full expressions and compare them to the Born-level and classical treatments, that we present in additional calculations. Making contact in these asymptotic regimes, gives credence to the correctness of our results.

We then study the kinematic regime of soft photon emission where the emitted photon energy is much smaller than the available center-of-mass energy. Here one may tap into the soft-photon theorems that imply the factorization of the cross section into the elastic cross section, multiplied by an emission piece. Weinberg recently showed how a formula for soft photon emission may be constructed for arbitrary $\nu_i$ for the case of dipole radiation. With some modifications, we are able to carry these ideas over to the case of quadrupole emission, and Eq.~\eqref{eqn:quad_emission_weinberg} gives the soft photon cross section that applies in both, the Born and the classical regime. Taking this approach further, we also show how an approximate formula valid for arbitrary $\nu_i$ and accross the kinematic range of $x$ can be constructed for repulsive potentials in~\eqref{eqn:quad_approximate}.

Equipped with all this knowledge on the quadrupole bremsstrahlung process, in a final section we then propose an adequate form for a Gaunt factor for electron-electron scattering which is based on practical demand. A numerical tabulation over a large kinematic regime will be presented in an upcoming work~\cite{inprep2020}. The emission of photons in the Coulomb collision of unpolarized free particles is perhaps {\em the} fundamental process in the interaction of light with matter. This work closes a seeming gap in the literature, by laying out the exact theory for quadrupole radiation in the non-relativistic limit and by studying its consequences in all relevant kinematic regimes.

  \paragraph*{Acknowledgements: }\ \  The authors are supported by the New Frontiers program
  of the Austrian Academy of Sciences. LS is supported by the Austrian
  Science Fund FWF under the Doctoral Program W1252-N27 Particles and
  Interactions. We acknowledge the use of computer packages for
  algebraic calculations~\citep{Mertig:1990an,Shtabovenko:2016sxi}.

\appendix

\section{Interference term in the full result}
\label{sec:coefficients}

The interference term $|I^{\alpha \beta}|^2 $ in (\ref{eqn:quad_tensor_identical_particles}) is composed as $|I^{\alpha \beta}|^2 =Q_1^{*\alpha \beta} Q_2^{\gamma \delta} +  Q_2^{*\alpha \beta} Q_1^{\gamma \delta}$ where $\alpha \beta = \gamma \delta =  a b$ or $\alpha \beta =  a a$ and $\gamma \delta =  b b$, respectively.  We find
\begin{align}\label{eqn:I_interference_app}
\frac{1}{2}|I^{\alpha \beta}|^2
		&=
		G_0 
		\Big\{
        A^{+}_{\thinhat\alpha\thinhat\beta}  \Re\left[F^*(z)F(\tilde z)\right]
		+ 
		A^{-}_{\thinhat\alpha\thinhat\beta}  \Im\left[F^*(z)F(\tilde z)\right] 
		+
		B^{+}_{\thinhat\alpha\thinhat\beta}\Re\left[F'^*(z)F'(\tilde z)\right]
		+
		B^{-}_{\thinhat\alpha\thinhat\beta} \Im\left[F'^*(z)F'(\tilde z)\right] 
		\nonumber\\
		&+
		C^{+}_{\thinhat\alpha\thinhat\beta}\Re\left[F^*(z)F'(\tilde z)\right] 
		+ 
		C^{-}_{\thinhat\alpha\thinhat\beta}\Im\left[F^*(z)F'(\tilde z)\right] 
        +
		D^{+}_{\thinhat\alpha\thinhat\beta}\Re\left[F'^*(z)F(\tilde z)\right]
		+ 
		D^{-}_{\thinhat\alpha\thinhat\beta}\Im\left[F'^*(z)F(\tilde z)\right]
			\Big\}
\end{align}
where $G_0$ is given by Eq.~\eqref{eqn:G_0}. For $\left| Q^{aa}\right|^2$  the coefficients read
	\begin{align}\label{eqn:Qnn_id}
		A^+_{\thinhat a\thinhat a} & =
		\frac{4   (y+1)^2}{(1-z) (1-\tilde z)}  \,,\qquad   
		B^+_{\thinhat a\thinhat a} = 
		\frac{9 y^2}{\nu_i^2}  \,, \qquad
		C^-_{\thinhat a\thinhat a} =
		 \frac{6  y (y+1)}{ \nu_i (z-1)}  \,, \qquad 
		D^-_{\thinhat a\thinhat a}  = 
		-\frac{6 y (y+1) }{\nu_i (\tilde z-1) }  \,,
	\end{align}
with $A^-_{\thinhat a\thinhat a}=B^-_{\thinhat a\thinhat a}=C^+_{\thinhat a\thinhat a}=D^+_{\thinhat a\thinhat a}=0$. For $| Q^{ab}|^2$ one gets the  lengthy expressions
{
\medmuskip=0mu
\thinmuskip=0mu
\thickmuskip=0mu
\nulldelimiterspace=1pt
\scriptspace=0pt
	\begin{align}
	A^+_{\thinhat a\thinhat b} &=
			\frac{4 \left(y^8+1\right)-2 y^2\left(y^4+1\right) \left(1+7 \cos^2\theta\right)+4 y^4 \left(1+2 \cos^4\theta+3 \cos^2\theta\right)}{(1-y)^6 (1-z)^2(1-\tilde z)^2}
		+
		\frac{2 \left(1-\cos^2\theta\right) \left[y^4+2y^2 \left(2 \cos^2\theta-3\right)+1\right]}{(1-y)^4 (1-z)^2(1-\tilde z)^2} \nu_i^2   ,
		\notag \\
	A^-_{\thinhat a\thinhat b} &=
	 -\frac{4 \nu_i   (y+1)^3 \cos \theta}{(y-1)^2\left(1-z\right)^2\left(1-\tilde z\right)^2} ,
		\quad
	B^+_{\thinhat a\thinhat b} = 
		\frac{y^2 \left[3 y^2-2 y \left(2+\cos^2\theta\right)+3\right]}{(1 - y)^2 \nu_i^2} 
		-
		\frac{8y^3 \left(1-\cos^2\theta\right)^2}{(1 - y)^4 (1-z)(1-\tilde z)} ,
		\quad
	B^-_{\thinhat a\thinhat b} = 
		\frac{4 y^3 (1+y) \cos\theta \left(1-\cos^2\theta\right)}{(1 - y)^4 (1-z)(1-\tilde z) \nu_i} ,
		\notag \\
	C^+_{\thinhat a\thinhat b} &=
		\frac{y \left(1+\cos\theta\right) \left[y^4-6 y^2+8 y^2 \cos^3\theta-4 \left(y^3+y^2+y\right)  \cos^2\theta+4 \left(-y^4+y^3+2 y^2+y-1\right) \cos\theta+1\right]}{(1-y)^4 (1-z)^2(1-\tilde z)} ,
		\notag \\ \label{eqn:Qmn_id}
	C^-_{\thinhat a\thinhat b} &= 
		\frac{2 y \left[(1-y) y \left(2 y^3 + y + 2\right) - 2\right] 
		- 2 y (y+1) \left[y \left((y-3)^2 y-6\right)+1\right]\cos\theta 
		+ 2 y^2 (y+1) \left[y (2 y-3)+2\right] \cos^2\theta
		-2 y^3 (y+1) \cos^3\theta
		}{(1-y)^4 (1-z)^2 \nu_i} 
		\notag\\
		& \quad +
		\frac{2 y (y+1) \left(\cos\theta-1\right) (1+\cos\theta)^2 \left[y \Big(y-4+2 \cos\theta\Big)+1\right]}{(1-y)^4 (1-z)^2(1-\tilde z)} \nu_i ,
		 \notag \\
	D^+_{\thinhat a\thinhat b} &= 
		\frac{y\left(1-6 y^2+y^4\right) 
		+ y \left[(y-2) y \Big(y (3 y+2)+2\Big)+3\right] \cos\theta
		-4 \left(y^5-y^3+y\right)\cos^2\theta
		+4 y^2 \left[(y-1) y+1\right]\cos^3\theta
		+8 y^3\cos^4\theta
		}{(1-y)^4 (1-z)(1-\tilde z)^2}\,,
		\notag \\
	D^-_{\thinhat a\thinhat b} &= 
		-\frac{
		2 y (y+1) \left\{(y^4+1) (\cos\theta-2)+2 y(y^2+1) (\cos\theta-2) (\cos\theta-1)+y^2 \left[\cos\theta \Big(9+(\cos\theta-3) \cos\theta\Big)-5\right]\right\}
		}{(1-y)^4 (1-\tilde z)^2 \nu_i} 
		\notag \\
		& \quad +
		\frac{2 y (y+1) (\cos\theta-1)^2 (1+\cos\theta) \left[y^2-2 y (2+\cos\theta)+1\right]}{(1-y)^4 (1-z)(1-\tilde z)^2} \nu_i\,.
		\end{align}
}%
Note that for identical particles $Z_1 Z_2 = Z^2 > 0$ and therefore in the coefficients in this section $\nu_i$ is always negative, as can be seen from Eq.~\eqref{eqn:nuif}.

\section{Born approximation}
\label{sec:mat_elements}

\begin{figure*}[tb]
\centering
\includegraphics[width=\textwidth]{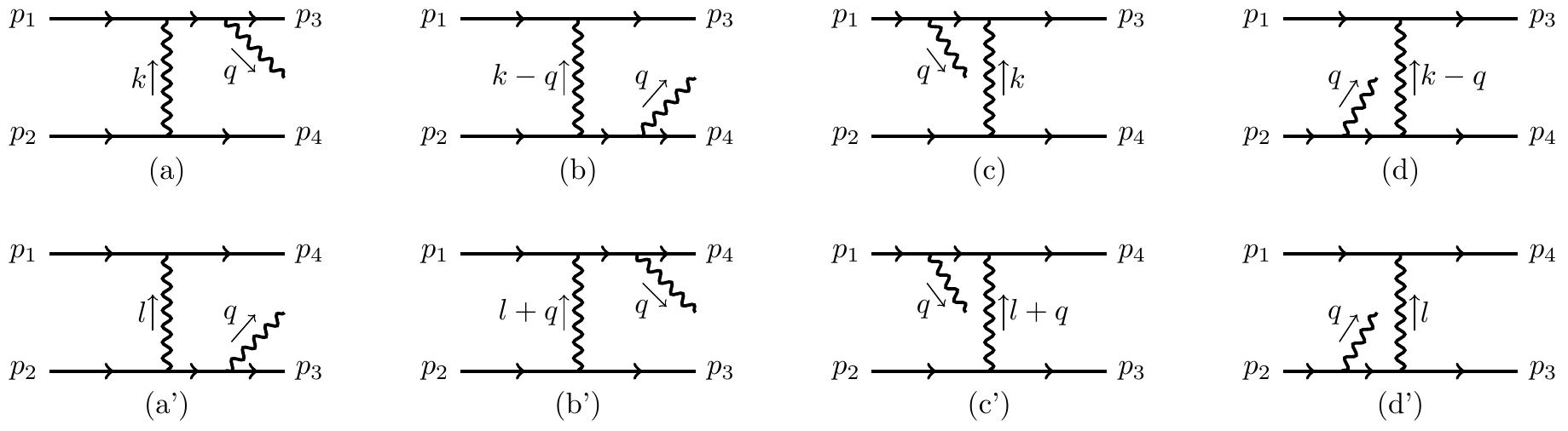}
\caption{Eight tree level diagrams contributing to $e^- e^-$-bremsstrahlung in the Born limit. Fermion momentum flows from left to right and the direction of the photon momentum is represented by the arrow next to the photon line. For $e^+ e^-$-bremsstrahlung only the non-primed diagrams have to be considered and the fermion flow is reversed for positrons.
		\label{fig:feyman_diag}}
              \end{figure*}

In the following, we present the  non-relativistic squared matrix elements in the Born approximation. For simplicity, we specialize to $m_1 = m_2$ and $|Z_{1,2}|=1$. The matrix elements can be obtained from a quantum field theory calculation, based on the QED Lagrangian, of the Feynman diagrams shown in Fig.~\ref{fig:feyman_diag}. The matrix elements are hence given in relativistic normalization $\langle \vec  p | \vec p' \rangle = (2\pi)^3 2E_{\vec p} \delta^3(\vec p - \vec p')$ 
and are summed over photon polarizations, averaged over initial and summed over final spins, and averaged over the direction of the emitted photon.
We stick to the notation of the main segment and use the variables $\vec k = \vec p_2 - \vec p_4$, $\vec l = \vec p_4 - \vec p_1$ and their directions $\hat k = \vec k/|\vec k|$, $\hat l = \vec l /|\vec l|$.

The scattering of $e^+e^-\to e^+e^-\gamma$ is dominated by dipole emission and only the unprimed diagrams on the left side of Fig.~\ref{fig:feyman_diag} contribute. To leading order in relative velocity of the colliding pair the squared matrix element reads
\begin{align} \label{eqn:matrix_dipole_ferm_vector}
	 \Big\langle\sum_{\lambda}	|\mathcal M^{\text{rel.}}_D |^2  \Big\rangle_{\hat q}&=
		\frac{128}{3} \frac{(4\pi\alpha)^3}{\omega^4} 
		 |\bm l|^2 (\hat {\bm k} \cdot \hat {\bm l} )^2\,.
	\end{align}
        In turn, for the scattering of two identical spin-1/2 (e.g.~$e^\pm e^{\pm}\to e^\pm e^\pm \gamma$) or spin-0 particles   (e.g.~$\alpha \alpha \to \alpha \alpha \gamma$), also the exchange diagrams on the right side of Fig.~\ref{fig:feyman_diag} contribute.  We obtain
\begin{align}
	 & \Big\langle\sum_{\lambda}	|\mathcal M^{\text{rel.}}_Q |^2  \Big\rangle_{\hat q}   = \frac{32}{15} \frac{(4\pi\alpha)^3}{\omega^2} 
|\bm k|^2 |\bm l|^2  \left( \frac{3 + 13(\hat {\bm k} \cdot \hat {\bm l} )^2}{|\bm k|^4} +\frac{3 + 13(\hat {\bm k} \cdot \hat {\bm l} )^2}{|\bm l|^4} 		
+ 2\frac{(-1)^{2s}}{2s+1}
		\frac{3 + 7(\hat {\bm k} \cdot \hat {\bm l} )^2 + 6(\hat {\bm k} \cdot \hat {\bm l} )^4}{|\bm k|^2 |\bm l|^2} 
\right) , \label{eqn:matrix_quadrupole_ferm_vector}
\end{align}
where $s$ is the spin of the colliding particles.
For $s=1/2$, Eq.~\eqref{eqn:matrix_quadrupole_ferm_vector} is in agreement with the original calculation by \cite{Fedyushin:1952hg} and for $s=0$ in agreement with~\cite{1981PhRvA..23.2851G}.
To obtain the matrix elements \eqref{eqn:matrix_dipole_ferm_vector} and \eqref{eqn:matrix_quadrupole_ferm_vector} we have used a velocity expansion in the relative velocity $v_i$ where $|\vec k |/\mu$ and $|\vec l |/\mu$ scale as $\mathcal O(v_i)$ and $(\vec k \cdot \vec l)/\mu^2$ and $\omega/\mu$ as $\mathcal O(v_i^2)$.

The cross sections in Sec.~\ref{sec:born_limit} can then be obtained from \eqref{eqn:matrix_dipole_ferm_vector} and \eqref{eqn:matrix_quadrupole_ferm_vector} by putting the corresponding matrix elements into Eq.~\eqref{eqn:diff_cross_section} after converting them to non-relativistic normalization, \textit{i.e.}
\begin{align}
    |\mathcal M_I|^2 = 2\omega (2m_1)^2(2m_2)^2   |\mathcal M^{\text{rel.}}_I|^2 \,,
\end{align}
and setting $|\vec k |/\mu=|\vec v_i - \vec v_f|$, $|\vec l |/\mu=|\vec v_i + \vec v_f|$ and $(\vec k \cdot \vec l)/\mu^2 = v_i^2-v_f^2$.

\section{Classical Energy Loss}
\label{sec:classical_cs}
The effective energy loss $\Eeffclass$ can be calculated semi-classically by integrating the intensity of the radiation over the impact parameter $\rho$ and the interaction time $t$ as (see e.g.~\cite{landau1975classical}) 
\begin{align}\label{eqn:classical_EL}
	\Eeffclass_I &= 
	\int_{0}^{\infty} d\rho \,2 \pi \rho \int_{-\infty}^{\infty} dt  \; I_I \,.
\end{align}
The intensity $I$ of the dipole or quadrupole radiation respectively is given by %
\begin{align}
	I_D &= \frac{1 }{6 \pi}  A_D^2|  \ddot{\bm d}(t) |^2,\qquad
	I_Q = \frac{1}{720 \pi} A_Q^2 |  \dddot{ q}^{a b}(t) |^2 \,,
\end{align}
with
\begin{align}
	\bm d(t) &= e \;\bm r (t), \qquad
	q^{ab}(t) = e \left[ 3 \,r^a(t) r^b(t) - r^2(t) \delta^{ab}\right]. \label{eq:quad_tensor_classical}
\end{align}
where $\vec r(t)$ is the particles' distance and $v = | \dot{\vec r}|$ will be the magnitude of the relative velocity to be used below. 
The time integral in Eq.~\eqref{eqn:classical_EL} can be replaced by an integral over the emitted energy $\omega$ of the Fourier modes of $\bm d$ and  $q^{ab}$, i.e.
\begin{align}
  \label{eq:EffClassical}
	\Eeffclass &= \frac{1}{4\pi}\int d\rho \, \rho \int d\omega \left[ \frac{2\omega^4}{3} A_D^2 |  {\tilde{\vec d}}(\omega) |^2 
	+  \frac{\omega^6}{180} A_Q^2 |  {\tilde{ q}}^{ab}(\omega) |^2\right] .
\end{align}
with $\ddot{\bm d}(t) = 1/(2\pi)\int_{-\infty}^{\infty} d\omega \,  \tilde {\vec d}(\omega)\, e^{-i \omega t}$ and $\dddot{q}^{ab}(t) = 1/(2\pi)\int_{-\infty}^{\infty} d\omega  \, \tilde{ q}^{ab}(\omega)\, e^{-i \omega t}$,
where we denote Fourier-transformed quantities with a tilde.
The particles' distance vector $\bm r(t)$ can be parametrized in a plane spanned by the orthonormal vectors $\hat e_1$ and $\hat e_2$ \citep{landau1975classical} 
\begin{subequations}\label{eqn:parametrization_classical_total}
\begin{align}\label{eqn:parametrization_classical}
	\vec r_1 &= \frac{\alpha |Z_1 Z_2|}{\mu v^2} (\epsilon \pm \cosh \xi) \, \hat e_1,  \\ 
	\vec r_2&= \frac{\alpha |Z_1 Z_2|}{\mu v^2} \sqrt{\epsilon^2-1} \sinh \xi \, \hat e_2, \\ 
	t&= \frac{\alpha |Z_1 Z_2|}{\mu v^3} (\epsilon \sinh \xi \pm \xi) \, .
	\label{eq:tofxi}
\end{align}
\end{subequations}
Here, $+(-)$ is to be used for repulsive  (attractive) interactions; $r_n=|\vec r_n|$ is parameterized by the eccentricity $\epsilon$ and~$\xi$. 
For unbound orbits, the eccentricity relates to the impact parameter as $\epsilon^2 = 1+\mu^2 \rho^2 v^4/(\alpha Z_1 Z_2)^2$ and we can rewrite the integral over $\rho$ with $0<\rho<\infty$  into an integral over $\epsilon$ with $1<\epsilon<\infty$, as we will do in the following.

\subsection{Dipole Emission}
For dipole emission in an attractive potential, the differential cross section with respect to the emitted energy is found from the first term in (\ref{eq:EffClassical}) by omitting the integration over $d\omega$, using the definition \eqref{eqn:Eeff}, and identifying $v$ with $v_i$~\citep{landau1975classical}
\begin{align}
\left.	\omega \frac{d\sigma_D}{d\omega}  \right|_\text{classical}
	&= \frac{4 \alpha^3 Z_1^2 Z_2^2 \omega^4 A_D^2}{3\mu^2 v_i^4} \int_1^\infty d\epsilon \;\epsilon \;\left(| \tilde r_1(\omega) |^2 + | \tilde r_2(\omega) |^2\right) \,,
	\label{eqn:dipole_xy}
\end{align}
where we have used that
\begin{subequations}
\begin{align}
	\tilde r_1(\omega) &= \int_{-\infty}^{\infty} dt \;  r_1 \; e^{i \omega t} 
	= \frac{\pi \alpha |Z_1 Z_2|}{\mu v_i^2\omega} H_{i \kappa}^{(1)\prime}(i\kappa\epsilon) \,, \\
	\tilde r_2(\omega) &= \int_{-\infty}^{\infty} dt  \; r_2 \; e^{i \omega t}  
	= -\frac{\pi \alpha |Z_1 Z_2|}{\mu v_i^2\omega} \frac{\sqrt{\epsilon^2-1}}{\epsilon} H_{i \kappa}^{(1)}(i\kappa\epsilon) \,,
\end{align}
\end{subequations}
with $\kappa = \alpha |Z_1 Z_2| \omega/ (\mu v_i^3)$; the integrals are solved by expressing $dt$ in terms of $d\xi$ using (\ref{eq:tofxi}). For dipole emission, a closed solution for the integral in Eq.~\eqref{eqn:dipole_xy} exist. Therefore the differential cross section for an attractive potential $Z_1Z_2 <0$ can be written as
\begin{align}\label{eqn:classical_cs_full}
		\left.\omega\frac{d\sigma_D}{d\omega} \right|_{\text{classical}}&=
		\frac{4 \pi^2}{3} \frac{\alpha^3 Z_1^2 Z_2^2 A_D^2}{\mu^2 v_i^2} \kappa
		\left| H_{i\kappa}^{(1)}\left(i \kappa\right ) \right| 
		H_{i\kappa}^{(1) \prime}\left(i \kappa\right ) ,
\end{align}
For a repulsive interaction, the respective sign in the parametrization in Eq.~\eqref{eqn:parametrization_classical_total} simply leads to an overall factor $\exp(-2 \pi \kappa)$  in Eq.~\eqref{eqn:classical_cs_full}. 

\subsection{Quadrupole Emission}
The cross section for quadrupole emission is found from the second term in (\ref{eq:EffClassical}) with the quadrupole tensor 
\eqref{eq:quad_tensor_classical}
and $r_n$ from \eqref{eqn:parametrization_classical_total}. The components of $ q^{ab}(t)$ are then Fourier-transformed to obtain $\tilde{q}^{ab}(\omega)$.  This is then plugged into \eqref{eq:EffClassical} to obtain the differential cross section (if the $\omega$ integration is omitted), which for attractive potentials $Z_1Z_2<0$ reads
\begin{align}
	\left. \omega \frac{d\sigma_Q}{d\omega}  \right|_{\text{classical}}
	&= \frac{\alpha^3 Z_1^2 Z_2^2 \omega^6 A_Q^2}{15\mu^2 v_i^4} \int_1^\infty d\epsilon \;\epsilon \; \Big[ \left| \tilde r_{11}(\omega) \right|^2 + \left| \tilde r_{22}(\omega) \right|^2 
	 + 3 \left|\tilde r_{12}(\omega)\right|^2
	-\Re\left(\tilde r_{11}(\omega) \; \tilde r_{22}^*(\omega)\right)
	\Big] .
\label{eqn:eqn:classical_quad_cs}
\end{align}
Similarly to the case of dipole emission, the quadrupole emission for a repulsive interaction $Z_1Z_2>0$ can be obtained from \eqref{eqn:eqn:classical_quad_cs} by multiplying it with an overall factor $\exp(-2 \pi \kappa)$. The coefficients read
\begin{align}
	\tilde r_{11}(\omega) &= \int_{-\infty}^{\infty} dt \;  r_1^2 \; e^{i \omega t} 
	 = 
	a
	\left[\frac{\epsilon^2-1}{\epsilon} \, \kappa \; H_{i \kappa}^{(1)\prime}(i\kappa \epsilon) -  \frac{i}{\epsilon^2} H_{i \kappa}^{(1)}(i\kappa\epsilon)\right] \,, \\
	\tilde r_{22}(\omega) &=\int_{-\infty}^{\infty} dt \;  r_2^2 \; e^{i \omega t} 
	 = 
	-a
	\left[\frac{\epsilon^2-1}{\epsilon}\kappa \;H_{i \kappa}^{(1)\prime}(i\kappa\epsilon) +  i \frac{\epsilon^2-1}{\epsilon^2} H_{i \kappa}^{(1)}(i\kappa\epsilon)\right] \,, 
	\\
	\tilde r_{12}(\omega) &=\int_{-\infty}^{\infty} dt \;  r_1 \; r_2\; e^{i \omega t} = 
	a\sqrt{\epsilon^2-1}
	\left[\frac{i}{\epsilon} \;H_{i \kappa}^{(1)\prime}(i\kappa\epsilon) -   \frac{\epsilon^2-1}{\epsilon^2} \kappa \; H_{i \kappa}^{(1)}(i\kappa\epsilon)\right] \,,
\end{align}
with $a=2\pi \alpha |Z_1 Z_2|/(\mu v_i \omega^2)$. 

\subsection{Asymptotic expressions for large \boldmath$\kappa$}
\label{sec:classical_Airy}

In the limit where $\kappa \gg 1$, which corresponds to the limit assumed in Sec.~\ref{sec:classical_limit}, we can approximate the Hankel functions in terms of Airy functions, \textit{i.e.}, 
\begin{align}
    H_{i \kappa}^{(1)}(i\kappa\epsilon)
    &\approx
    \frac{ 2^{4/3} }{ i (\epsilon \kappa)^{1/3} } \mathrm{Ai}\left(\frac{2^{1/3}(\epsilon -1)\kappa}{(\epsilon \kappa)^{1/3}}\right) \,,
    \\
    H_{i \kappa}^{(1)\prime}(i\kappa\epsilon)
    &\approx
    -\frac{ 2^{5/3} }{ (\epsilon \kappa)^{2/3} } \mathrm{Ai}'\left(\frac{2^{1/3}(\epsilon -1)\kappa}{(\epsilon \kappa)^{1/3}}\right) \,.
\end{align}
Putting these expressions into \eqref{eqn:dipole_xy} or \eqref{eqn:eqn:classical_quad_cs} respectively, it can be shown that for $\kappa \to \infty$ the integrand goes to zero everywhere except at $\epsilon=1$. Thus, expanding the argument of the Airy functions to leading order and the prefactors to fourth order in $(\epsilon-1)$, one obtains the asymptotic expressions in $\kappa$, \eqref{sigma_dipole_large_nu_att} and \eqref{sigma_quad_large_nu_att}, which in Sec.~\ref{sec:classical_limit} were obtained from the exact quantum mechanical cross sections.

\section{Approximate Formula for Arbitrary~$\nu_i$}
\label{app:approximate_formula}

\begin{figure*}[t]
    \centering
	\includegraphics[width=\textwidth]{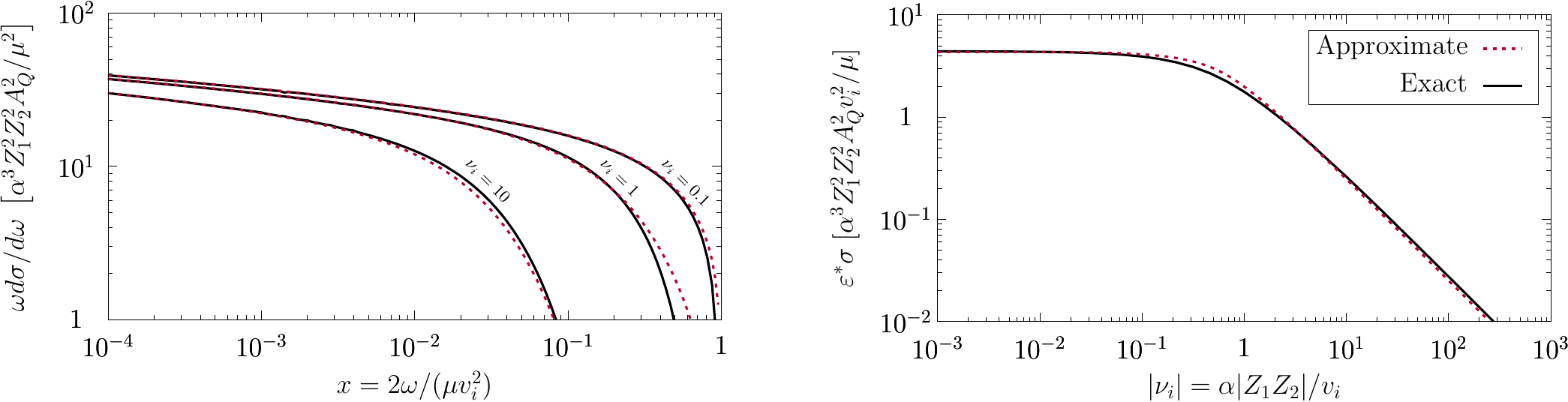}%
	\caption{Approximate formula \eqref{eqn:quad_approximate} for non-identical particles compared to the exact result \eqref{eqn:dist_cross_section}. The differential cross section is shown for $\nu_i=0.1,1,10$ in the top panel and the effective energy loss in the bottom panel. The equivalent approximate result for identical particles is {\em on par} in terms of accuracy w.r.t.~the exact result.
	\label{fig:approximate_formula}}
\end{figure*}

For repulsive interactions, and building on our insights on the process in the various kinematic regimes, one may in fact go beyond Eq.~\eqref{eqn:quad_emission_weinberg} and find a ``doctored'' formula that is not only valid for arbitrary $\nu_i$ but also extends the validity away from the soft-photon regime. Although such formula cannot be derived from first principles, it can nevertheless be helpful to obtain reasonably accurate results for quadrupole emission across the entire region in $x$ and $\nu_i$ from a simple expression. 

For this, we first multiply Eq.~\eqref{eqn:quad_emission_weinberg} by the exponential factor $\exp(-2\pi\kappa)$ that is ubiquitous in the classical formul\ae\ for repulsive interactions. Then we supply  a  phase space factor $\sqrt{1-x}$ which was neglected in the soft limit and take the $x$-dependent term in the logarithm to be the one from the full Born cross section. In addition, we add to this the classical cross section for $\kappa\gg1$ to cover hard photon emission. The resulting approximate formula is
\begin{align} \label{eqn:quad_approximate}
    \omega \frac{d \sigma_{Q}}{d\omega} &\approx
    \frac{4}{15} \frac{\alpha^{3}Z_1^2 Z_2^2 A_Q^2}{\mu^2} \frac{\sqrt{1-x}}{e^{\pi x |\nu_i|}}
    \Bigg\{
    \frac{40 \pi^{3/2}}{3^{1/6} \Gamma(1/6)} \left(\frac{x|\nu_i|}{2}\right)^{2/3}
    + \frac{\pi^2 \nu_i^2}{\sinh^2{\pi \nu_i}}\left(6\frac{2+\xi}{\zeta^2+1}-3(2+\xi)+ 12\ln \zeta
    \right) 
    \nonumber \\
    &+26 
    -6\frac{2+\xi}{\zeta^2+1}
    +12\ln \left(\frac{1}{\zeta}\frac{1+\sqrt{1-x}}{1-\sqrt{1-x}}\right) 
    \Bigg\} 
    ,
\end{align}
with $\zeta=|\nu_i| e^{\gamma+1/2}$ and $\xi=(0)1$ for (non-)identical particles. This approximate cross section is easy to compute numerically in contrast to the exact results \eqref{eqn:dist_cross_section} and \eqref{eq:quad_cs_full_id} and yields reasonable results in all regimes.

Equation~\eqref{eqn:quad_approximate} is essentially exact for arbitrary $\nu_i$ for soft photons, as shown in the top panel in Fig.~\ref{fig:approximate_formula}. At the kinematic endpoint, for $x\to 1$, it deviates from the exact result. In the Born limit $|\nu_i|\ll 1$, a deviation is only observed very close to $x=1$ where the Born cross section is already kinematically suppressed. In the classical limit, $|\nu_i|\gg 1$, the deviation is visible for smaller values of $x$, but in a region where the cross section is already highly Coulomb suppressed. For this reason, Eq.~\eqref{eqn:quad_approximate} turns out to be an excellent approximation for the effective energy loss demonstrated in the bottom panel of Fig.~\ref{fig:approximate_formula}. The effective energy loss deduced from Eq.~\eqref{eqn:quad_approximate} has an error of roughly 1\% w.r.t~the exact result in the Born regime, in the intermediate regime $\nu_i \sim 1$ the error reaches its maximum of 15\% and in the classical regime the exact result is underestimated by 10\%.

\bibliography{references.bib}

\begin{thebibliography}{}
\expandafter\ifx\csname natexlab\endcsname\relax\def\natexlab#1{#1}\fi
\providecommand{\url}[1]{\href{#1}{#1}}
\providecommand{\dodoi}[1]{doi:~\href{http://doi.org/#1}{\nolinkurl{#1}}}
\providecommand{\doeprint}[1]{\href{http://ascl.net/#1}{\nolinkurl{http://ascl.net/#1}}}
\providecommand{\doarXiv}[1]{\href{https://arxiv.org/abs/#1}{\nolinkurl{https://arxiv.org/abs/#1}}}

\bibitem[{Abramowitz \& Stegun(1948)}]{abramowitz1948handbook}
Abramowitz, M., \& Stegun, I.~A. 1948, Handbook of mathematical functions with
  formulas, graphs, and mathematical tables, Vol.~55 (US Government printing
  office)

\bibitem[{Akhiezer \& Berestetskii(1953)}]{akhiezer1953quantum}
Akhiezer, A.~I., \& Berestetskii, V.~B. 1953, Quantum electrodynamics, Vol.
  2876 (United States Atomic Energy Commission, Technical Information Service
  Extension)

\bibitem[{Berestetskii {et~al.}(1982)Berestetskii, Lifshitz, \&
  Pitaevskii}]{berestetskii1982quantum}
Berestetskii, V., Lifshitz, E., \& Pitaevskii, L. 1982, Quantum
  Electrodynamics, Course of theoretical physics (Butterworth-Heinemann)

\bibitem[{Chluba {et~al.}(2020)Chluba, Ravenni, \& Bolliet}]{Chluba:2019ser}
Chluba, J., Ravenni, A., \& Bolliet, B. 2020, Mon. Not. Roy. Astron. Soc., 492,
  177, \dodoi{10.1093/mnras/stz3389}

\bibitem[{{Dermer}(1984)}]{1984ApJ...280..328D}
{Dermer}, C.~D. 1984, \apj, 280, 328, \dodoi{10.1086/161999}

\bibitem[{Elwert(1939)}]{Elwert:1939km}
Elwert, G. 1939, Ann. Phys., 426, 178.
\newblock \url{https://doi.org/10.1002/andp.19394260206}

\bibitem[{Fedyushin(1952)}]{Fedyushin:1952hg}
Fedyushin, B. 1952, Zh. Eksperim. Teor. Fiz., 22, 140

\bibitem[{{Felten} {et~al.}(1966){Felten}, {Gould}, {Stein}, \&
  {Woolf}}]{1966ApJ...146..955F}
{Felten}, J.~E., {Gould}, R.~J., {Stein}, W.~A., \& {Woolf}, N.~J. 1966, \apj,
  146, 955, \dodoi{10.1086/148972}

\bibitem[{Gal'stov \& Grats(1976)}]{Galstov1976}
Gal'stov, D.~V., \& Grats, Y.~V. 1976, Theoretical and Mathematical Physics,
  28, 730, \dodoi{10.1007/bf01029030}

\bibitem[{Gal'tsov \& Grats(1974)}]{Galtsov1974}
Gal'tsov, D.~V., \& Grats, Y.~V. 1974, Soviet Physics Journal, 17, 1713,
  \dodoi{10.1007/bf00892884}

\bibitem[{{Gould}(1981)}]{1981PhRvA..23.2851G}
{Gould}, R.~J. 1981, \pra, 23, 2851, \dodoi{10.1103/PhysRevA.23.2851}

\bibitem[{{Gould}(1990)}]{1990ApJ...362..284G}
---. 1990, \apj, 362, 284, \dodoi{10.1086/169265}

\bibitem[{Gould(2006)}]{gould2006electromagnetic}
Gould, R.~J. 2006, Electromagnetic Processes, Princeton Series in Astrophysics
  (Princeton University Press)

\bibitem[{Haug \& Nakel(2004)}]{Haug:2004gp}
Haug, E., \& Nakel, W. 2004, {The elementary process of bremsstrahlung}

\bibitem[{{Hummer}(1988)}]{1988ApJ...327..477H}
{Hummer}, D.~G. 1988, \apj, 327, 477, \dodoi{10.1086/166210}

\bibitem[{{Itoh} {et~al.}(2002){Itoh}, {Kawana}, \&
  {Nozawa}}]{2002NCimB.117..359I}
{Itoh}, N., {Kawana}, Y., \& {Nozawa}, S. 2002, Nuovo Cimento B Serie, 117,
  359.
\newblock \doarXiv{astro-ph/0111040}

\bibitem[{{Johnson}(1972)}]{1972ApJ...174..227J}
{Johnson}, L.~C. 1972, \apj, 174, 227, \dodoi{10.1086/151486}

\bibitem[{Karzas \& Latter(1961)}]{1961ApJS....6..167K}
Karzas, W.~J., \& Latter, R. 1961, The Astrophysical Journal Supplement Series,
  6, 167, \dodoi{10.1086/190063}

\bibitem[{{Kellogg} {et~al.}(1975){Kellogg}, {Baldwin}, \&
  {Koch}}]{1975ApJ...199..299K}
{Kellogg}, E., {Baldwin}, J.~R., \& {Koch}, D. 1975, \apj, 199, 299,
  \dodoi{10.1086/153692}

\bibitem[{Kramers(1923)}]{Kramers:1923}
Kramers, H.~A. 1923, The London, Edinburgh, and Dublin Philosophical Magazine
  and Journal of Science, 46, 836, \dodoi{10.1080/14786442308565244}

\bibitem[{Landau \& Lifshitz(1975)}]{landau1975classical}
Landau, L., \& Lifshitz, E. 1975, The Classical Theory of Fields, Course of
  theoretical physics (Butterworth-Heinemann)

\bibitem[{Landau \& Lifshitz(1977)}]{landau1977quantum}
---. 1977, Quantum Mechanics: Non-relativistic Theory, Butterworth-Heinemann
  (Butterworth-Heinemann)

\bibitem[{Landau \& Pomeranchuk(1953)}]{Landau:1953um}
Landau, L., \& Pomeranchuk, I. 1953, Dokl. Akad. Nauk Ser. Fiz., 92, 535

\bibitem[{Low(1958)}]{Low:1958sn}
Low, F. 1958, Phys. Rev., 110, 974, \dodoi{10.1103/PhysRev.110.974}

\bibitem[{Maxon \& Corman(1967)}]{Maxon1967}
Maxon, M.~S., \& Corman, E.~G. 1967, Physical Review, 163, 156,
  \dodoi{10.1103/physrev.163.156}

\bibitem[{{Menzel} \& {Pekeris}(1935)}]{1935MNRAS..96...77M}
{Menzel}, D.~H., \& {Pekeris}, C.~L. 1935, \mnras, 96, 77,
  \dodoi{10.1093/mnras/96.1.77}

\bibitem[{Mertig {et~al.}(1991)Mertig, Bohm, \& Denner}]{Mertig:1990an}
Mertig, R., Bohm, M., \& Denner, A. 1991, Comput. Phys. Commun., 64, 345,
  \dodoi{10.1016/0010-4655(91)90130-D}

\bibitem[{Migdal(1956)}]{Migdal:1956tc}
Migdal, A. 1956, Phys. Rev., 103, 1811, \dodoi{10.1103/PhysRev.103.1811}

\bibitem[{Nordsieck(1954)}]{Nordsieck:1954hg}
Nordsieck, A. 1954, Phys. Rev., 93, 785.
\newblock \url{https://doi.org/10.1103/PhysRev.93.785}

\bibitem[{Oppenheimer(1929)}]{Oppenheimer1929}
Oppenheimer, J.~R. 1929, Z. Phys., 55, 725, \dodoi{10.1007/BF01330752}

\bibitem[{Pradler \& Semmelrock(in preparation)}]{inprep2020}
Pradler, J., \& Semmelrock, L. in preparation

\bibitem[{Raffelt(1996)}]{Raffelt:1996wa}
Raffelt, G. 1996, {Stars as laboratories for fundamental physics}: {The
  astrophysics of neutrinos, axions, and other weakly interacting particles}

\bibitem[{Sarazin(1986)}]{Sarazin:1986zz}
Sarazin, C.~L. 1986, Rev. Mod. Phys., 58, 1, \dodoi{10.1103/RevModPhys.58.1}

\bibitem[{Shtabovenko {et~al.}(2016)Shtabovenko, Mertig, \&
  Orellana}]{Shtabovenko:2016sxi}
Shtabovenko, V., Mertig, R., \& Orellana, F. 2016, Comput. Phys. Commun., 207,
  432, \dodoi{10.1016/j.cpc.2016.06.008}

\bibitem[{Sommerfeld(1931)}]{Sommerfeld1931}
Sommerfeld, A. 1931, Ann. Phys., 403, 257, \dodoi{10.1002/andp.19314030302}

\bibitem[{Sommerfeld(1939)}]{Sommerfeld_Spektrallinien2}
---. 1939, Atombau und Spektrallinien II. (Friedr.~Vieweg \& Sohn,
  Braunschweig)

\bibitem[{Sommerfeld \& Maue(1935)}]{Sommerfeld:1935ab}
Sommerfeld, A., \& Maue, A. 1935, Ann. Phys., 415, 589.
\newblock \url{https://doi.org/10.1002/andp.19354150702}

\bibitem[{Sugiura(1929)}]{Sugiura1929}
Sugiura, Y. 1929, Phys. Rev., 34, 858, \dodoi{10.1103/PhysRev.34.858}

\bibitem[{{Terzian}(1965)}]{1965ApJ...142..135T}
{Terzian}, Y. 1965, \apj, 142, 135, \dodoi{10.1086/148268}

\bibitem[{van Hoof {et~al.}(2014)van Hoof, Williams, Volk, Chatzikos, Ferland,
  Lykins, Porter, \& Wang}]{vanHoof:2014bha}
van Hoof, P., Williams, R., Volk, K., {et~al.} 2014, Mon. Not. Roy. Astron.
  Soc., 444, 420, \dodoi{10.1093/mnras/stu1438}

\bibitem[{{Vocks} {et~al.}(2018){Vocks}, {Mann}, {Breitling}, {Bisi},
  {D{\k{a}}browski}, {Fallows}, {Gallagher}, {Krankowski}, {Magdaleni{\'c}},
  {Marqu{\'e}}, {Morosan}, \& {Rucker}}]{2018A&A...614A..54V}
{Vocks}, C., {Mann}, G., {Breitling}, F., {et~al.} 2018, \aap, 614, A54,
  \dodoi{10.1051/0004-6361/201630067}

\bibitem[{Weinberg(1965{\natexlab{a}})}]{Weinberg:1965nx}
Weinberg, S. 1965{\natexlab{a}}, Phys. Rev., 140, B516,
  \dodoi{10.1103/PhysRev.140.B516}

\bibitem[{Weinberg(1965{\natexlab{b}})}]{Weinberg:1965rz}
---. 1965{\natexlab{b}}, Phys. Rev., 138, B988,
  \dodoi{10.1103/PhysRev.138.B988}

\bibitem[{Weinberg(2015)}]{weinberg2015lectures}
---. 2015, Lectures on quantum mechanics (Cambridge University Press)

\bibitem[{Weinberg(2019)}]{Weinberg:2019mai}
---. 2019, Phys. Rev. D, 99, 076018, \dodoi{10.1103/PhysRevD.99.076018}

\end{thebibliography}
\end{document}